\documentclass[preprint,pteplogo,amsmath,amssymb]{ptephy}

\preprintnumber{KYUSHU-RCAPP-2020-01}

\usepackage{graphicx} 
\usepackage{dcolumn}  
\usepackage{bm}       
\usepackage{subfiles} 
\usepackage{natbib}   
\usepackage{wasysym}  

\newcommand{\Hecapture}{$^3$He(n,p)$^3$H}

\newcommand{\Ncapture}{$^{14}$N(n,p)$^{14}$C}
\newcommand{\Ocapture}{$^{17}$O(n,$\alpha$)$^{14}$C}
\newcommand{\Ccapture}{$^{12}$C(n,$\gamma$)$^{13}$C}
\newcommand{\Licapture}{$^6$Li(n,$\alpha$)$^3$H}

\newcommand{\LiF}{$^6$LiF tile}

\enlargethispage{2cm} 

\begin{document}
\title{Neutron lifetime measurement with pulsed cold neutrons}
\author{\name{K.~Hirota}{1}, \name{G.~Ichikawa}{2,3}, \name{S.~Ieki}{4}\footnote{Present address : Research Center for Neutrino Science, Tohoku University}, \name{T.~Ino}{2,3}, \name{Y.~Iwashita}{5},\name{M.~Kitaguchi}{1,6}, \name{R.~Kitahara}{7}\footnote{Present address : Sumitomo Heavy Industries, Ltd.}, \name{J.~Koga}{8}, \name{K.~Mishima\footnote{Corresponding author}}{2,3*}, \name{T.~Mogi}{4}, \name{K.~Morikawa}{1}, \name{A.~Morishita}{8}, \name{N.~Nagakura}{4}, \name{H.~Oide}{4}\footnote{Present address : Department of Physics, Tokyo Institute of Technology}, \name{H.~Okabe}{1}, \name{H.~Otono}{9}, \name{Y.~Seki}{10}, \name{D.~Sekiba}{11}, \name{T.~Shima}{12}, \name{H.~M.~Shimizu}{1}, \name{N.~Sumi}{2,3,8}, \name{H.~Sumino}{13}, \name{T.~Tomita}{8}\footnote{Present address : Tokyo Tatemono Co.,Ltd.}, \name{H.~Uehara}{8}, \name{T.~Yamada}{4}, \name{S.~Yamashita}{14}, \name{K.~Yano}{8}, \name{M.~Yokohashi}{1}, and \name{T.~Yoshioka}{9}}

\address{
\affil{1}{Department of Physics, Nagoya University, Nagoya, 464-8602, Japan}
\affil{2}{High Energy Accelerator Research Organization (KEK), Tsukuba, 305-0802, Japan}
\affil{3}{J-PARC Center, Tokai, 319-1195, Japan}
\affil{4}{Department of Physics, The University of Tokyo, Tokyo 113-0033, Japan}
\affil{5}{Institute of Chemical Research, Kyoto University, Uji, 611-0011, Japan}
\affil{6}{Kobayashi-Maskawa Institute for the Origin of Particles and the Universe (KMI), Nagoya University, Nagoya, 464-8602, Japan}
\affil{7}{Department of Physics, Kyoto University, Kyoto, 606-8502, Japan}
\affil{8}{Department of Physics, Graduate School of Science, Kyushu University, Fukuoka, 819-0395, Japan}
\affil{9}{Research Center for Advanced Particle Physics (RCAPP), Kyushu University, Fukuoka, 819-0395, Japan}
\affil{10}{Center for Physics and Mathematics, Osaka Electro-Communication University, Neyagawa, 572-8530, Japan}
\affil{11}{Institute of Applied Physics, University of Tsukuba, Tsukuba, 305-8573, Japan}
\affil{12}{Research Center for Nuclear Physics (RCNP), Osaka University, Ibaraki, 567-0047, Japan}
\affil{13}{Department of General Systems Studies, Graduate School of Arts and Sciences, The University of Tokyo, Tokyo, 153-8902, Japan}
\affil{14}{International Center for the Elementary Particle Physics (ICEPP), The University of Tokyo, Tokyo, 113-0033, Japan}
\email{kenji.mishima@kek.jp}}

\date{\today}
\begin{abstract}
The neutron lifetime has been measured by comparing the decay rate with the reaction rate of $^3$He nuclei of a pulsed neutron beam from the spallation neutron source at the Japan Proton Accelerator Research Complex (J-PARC).
The decay rate and the reaction rate were determined by simultaneously detecting electrons from the neutron decay and protons from the {\Hecapture} reaction using a gas chamber of which working gas contains diluted $^3$He.
The measured neutron lifetime was $898\,\pm\,10\,_{\rm stat}\,^{+15}_{-18}\,_{\rm sys}\,$s.
\end{abstract}
\subjectindex{C02, C30, D02, D40}

\maketitle
\section{Introduction}\label{sec_introduction}
A neutron decays into a proton, an electron, and an antineutrino through the weak interaction.
The decay lifetime is an important parameter for both cosmology and elementary particle physics.
The Big Bang Nucleosynthesis (BBN) is considered to create light elements, and the comparison of the observational data and the theoretical prediction for light element abundances provides a good opportunity to test cosmological models~\cite{tanabashi2018PDG,izotov2013primordial,aver2015effects,adam2016planck}. The neutron lifetime determines the number ratio of protons to neutrons at the beginning of the BBN, which affects the BBN yields of light elements, especially $^4$He~\cite{mathews2005big}.

In the Standard Model of particle physics, the neutron lifetime is described with a matrix element of $V_{\rm ud}$ in the Cabibbo-Kobayashi-Maskawa matrix. The neutron lifetime and the ratio of the weak axial-vector to vector coupling constants make it possible to determine the $V_{\rm ud}$~\cite{tanabashi2018PDG,mund2013determination, mendenhall2013precision,darius2017measurement,Markisch2019}.
The neutron lifetime is also demanded in the calculation of the cross section of the antineutrino capture reaction by a proton, which is the inverse reaction of the neutron beta decay~\cite{mention2011reactor}.

The neutron lifetime has been measured by many groups over the past fifty years~\cite{wietfeldt2011}.
The recent measurements were performed by two different experimental methods.
One is a so-called bottle method; the number of the surviving ultra-cold neutrons (UCNs) contained in a storage bottle is measured as a function of the elapsed time, and the lifetime is determined by fitting the data with an exponential decay curve~\cite{Mampe1993an, pichlmaier2010neutron, steyerl2012quasielastic, Arzumanov2015tea, serebrov2017new, pattie2018measurement, ezhov2018measurement}.
On the other hand, the beam method determines the neutron lifetime from the decay probability of the neutron obtained from the measured ratio of the decay rate to the incident neutron flux~\cite{Byrne1996zz,yue2013improved}.
The averaged neutron lifetimes are 879.4\,$\pm$\,0.4\,s and 888.0\,$\pm$\,2.0\,s for the bottle method
and the beam method, respectively. The central values differ by 8.5\,s, corresponding to the deviation of 4.0\,$\sigma$ using quoted uncertainties. 

The discrepancy is called the ``neutron lifetime puzzle'', and it is still unsettled whether it is due to any unconsidered systematic effect or any new physics.
As a solution for the neutron lifetime puzzle, several scenarios of exotic decay modes of neutron have recently been discussed. 
If a neutron decays into some undetectable particles with a branching ratio of about 1\%, for example, a mirror neutron~\cite{serebrov2008experimental} or dark particles~\cite{fornal2018dark}, the puzzle can be solved.
Note that some models with dark particles were already excluded~\cite{tang2018search,sun2018search, PhysRevLett.122.222503}
 and the characteristics of the dark particles are restricted by the astronomical data on massive neutron stars~\cite{McKeen2018,Motta2018,Baym2018, Grinstein2019}.

In the current situation, it is important to verify the puzzle by experiments in which different systematic errors dominate. We performed a new experiment with the beam method; the neutron lifetime was measured by the counting rate of the decay electrons relative to the {\Hecapture} reaction rate in a $^3$He-diluted gas detector. A great benefit is that this method is free from some systematic uncertainties thanks to the simultaneous measurement of the neutron flux and the neutron decay in the same detector volume, in contrast to the conventional beam methods which counted the decay protons~\cite{yue2013improved,nico2005measurement}.
It should be mentioned that our experiment measures the decay electrons but not decay protons, therefore, it has a sensitivity to the decay mode with no proton emission which is discussed in Ref.~\cite{fornal2018dark}.

This method was originally developed by Kossakowski $\it{et~al.}$~\cite{kossakowski1989neutron}. In their experiment, the diffracted neutron beam from a nuclear reactor was chopped into monochromatized bunches in order to separate the $\gamma$-ray background induced by neutron capture reactions on transmission through detector windows and the beam catcher. Our experiment was performed with the high-intensity pulsed neutron beam provided at the Japan Proton Accelerator Research Complex (J-PARC), which enables one to deliver such neutron bunches without loss due to monochromatization.

\section{Experiment}\label{sec_setup}
\subsection{Principle}\label{sec_principle}
In this experiment, electrons from the neutron decays are counted by observing the ionization tracks induced in the gas of a time projection chamber (TPC), because it is sensitive to electrons but not to $\gamma$-rays.
A thin $^{3}$He gas (50--200\,mPa) was admixed in the working gas 
in order to simultaneously measure the neutron flux by counting protons of 572 keV and tritons of 191 keV from the {\Hecapture} reactions.
The neutron lifetime, $\tau_{\rm n}$,  can be expressed as follows~\cite{kossakowski1989neutron}, 
\begin{equation}
\tau_{\rm n} = \frac{1}{\rho\sigma_0 v_0}\left( \frac{S_{\rm He}/\varepsilon_{\rm He}}{S_\beta/\varepsilon_\beta} \right),
\label{eq_tau_0}
\end{equation}
where $S_{\rm He}$ and $S_{\beta}$ are the numbers of observed events of the {\Hecapture} reactions and the decay electrons, respectively; $\varepsilon_{\rm He}$ and $\varepsilon_{\beta}$ are the detection efficiency of each reaction; $\rho$ is the number density of the $^3$He nuclei in the TPC.
Since the neutron absorption cross section is inversely proportional to the neutron velocity at low energies (known as the $1/v$ law), the product of the cross section and the velocity is constant.
Therefore, we can represent the reaction rate as $\sigma_{0}v_{0}$, where $\sigma_0$ is the cross section of the {\Hecapture} reaction, known as $5333\,\pm\,7$\,barn~\cite{Mughabghab2006} at the thermal neutron velocity of $v_0=2200$\,m/s.
The number density, $\rho$, is controlled by diluting the $^3$He gas at the calibrated conditions of volume, pressure, and temperature.
The efficiencies, $\varepsilon_{\rm He}$ and $\varepsilon_{\beta}$ in Eq.~(\ref{eq_tau_0}), are evaluated by Monte Carlo simulations which reproduce the responses of the TPC with sufficient accuracy.

\begin{figure}
	\begin{center}
	\includegraphics[width=0.7\columnwidth]{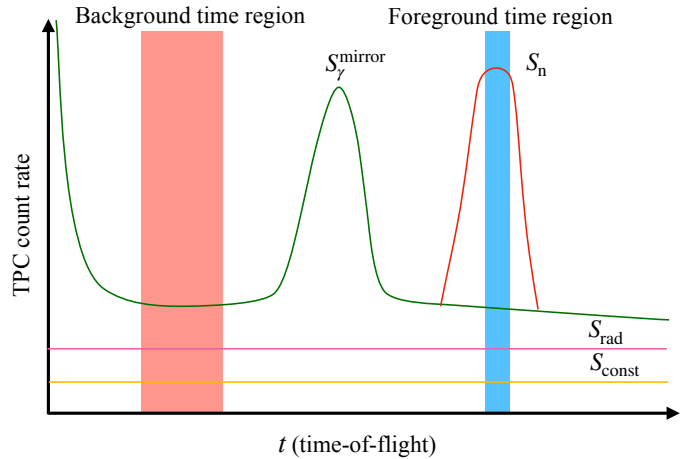}
	\caption{
	Schematic illustration of the distribution of charged particles reaching or produced in the sensitive region of the TPC as a function of time-of-flight, in case of a single bunch per pulse. 
	}
	\label{fig:TOF}
	\end{center}
\end{figure}

The numbers of events, $S_{\rm He}$ and $S_{\beta}$, are obtained by analyzing detected events in the TPC.
The signal and possible background events in this experiment are schematically shown in Fig.~\ref{fig:TOF}. 
Since events caused by neutrons occur when the neutrons are inside the TPC, they make a peak structure on the time-of-flight, $t$, and the number of events in the peak is denoted as $S_{\rm n}$.
The TPC detects background events by cosmic-rays or natural radiations. These $t$-independent backgrounds is denoted as $S_{\rm const}$. Events caused by neutrons, which is $t$-dependent, can be extracted by subtracting $S_{\rm const}$ by using the neutron-free region on $t$.
Neutron capture reactions at the neutron mirrors during the beam transport produce $\gamma$-rays. The number of backgrounds caused by the $\gamma$-rays is denoted as $S_{\gamma}^{\rm mirror}$. Because this background is $t$-dependent, it is evaluated by switching the beam to the TPC on and off using a neutron shutter. The neutron captures also create radioactive isotopes, and we denote the number of backgrounds coming from them as $S_{\rm rad}$. It depends on the lifetimes of the radioactive isotopes. 
If their lifetimes are longer enough than the period of the shutter-switching, their events are subtracted as well as $S_{\rm const}$. Thus the radioactive isotopes with short lives only appear when the shutter is open. Subtraction with/without the beam on $t$-regions and with open/closed of the shutter is applied to derive $S_{\rm n}$, which consists of $S_{\rm He}$, $S_{\beta}$, and other background events caused by the TPC working gas. Finally, $S_{\rm He}$ and $S_{\beta}$ are derived by applying some cuts and corrections to $S_{\rm n}$. Note that the $S$'s are defined as the number of events by each component in the foreground time region with the beam shutter open.

The experimental apparatus and procedure of the measurements are described in the rest of Sec.~\ref{sec_setup}, and the analysis is described in Sec.~\ref{sec_ana}.

\subsection{Neutron source and beamline}
A spallation neutron source at the Materials and Life Science Experimental Facility (MLF) in the J-PARC produces pulsed neutron beams by using 3\,GeV protons with a repetition rate of 25\,Hz. 
The neutron source emits fast neutrons on the injection of the primary proton beam, and the timing is defined as zero on $t$.
The neutrons are cooled down with liquid hydrogen moderators and transported to beamlines at the experimental halls of MLF. This experiment is conducted at ``Polarized-beam branch'' of the beamline BL05 (NOP)~\cite{mishima2009design}.
A schematic view of the beamline and experimental apparatus is illustrated in Fig.~\ref{fig_bl05setup}~\cite{arimoto2015development}. 
Neutrons are transported from a moderator to the experimental area through a polarizing neutron bender of $3.5\,$m filled with He gas, and vacuum guides of $4\,$m.
The time-averaged beam intensity at the exit of the vacuum guide (E in Fig.~\ref{fig_bl05setup}) corresponds to $(4.0\,\pm\,0.3)\,\times\,10^7$ s$^{-1}$\,cm$^{-2}$ at 1\,MW operation~\cite{nakajima2017materials,mishima2015neutron} with the beam polarization of 97--94\% in the wavelength of 0.2--0.9\,nm~\cite{ino2011measurement}.
The coordinate system used in this paper is depicted in the figure; 
the $z$-axis is in the beam direction at the TPC, $y$-axis is the vertical upward axis, and the $x$-axis is perpendicular to these so as to form a right-handed frame.

\subsection{Devices for the beam transport} \label{sec_beamtransport}
\begin{figure*}
\centering
  \includegraphics[width=0.95\columnwidth]{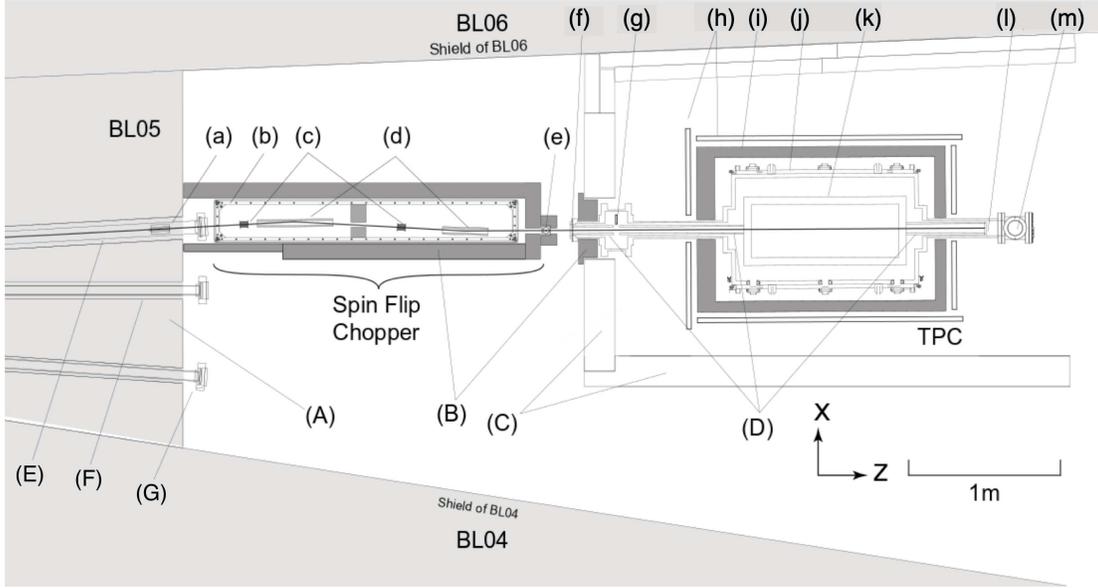}
  \caption{Schematic top view of experimental apparatus installed at Polarized neutron beam branch of NOP beamline: (A) concrete shield, (B) lead shields, (C) iron shield, (D) $^6$LiF beam collimator, (E) Polarized-beam branch, (F) Unpolarized-beam branch, (G) Low-divergence branch, (a) short-wavelength pass filter, (b) guide coil, (c) resonance spin flipper coils, (d) magnetic super mirrors, (e) neutron beam monitor, (f) 50-$\rm \mu$m-thick Zr window, (g) neutron switching shutter, (h) cosmic-ray veto counters, (i) lead shield, (j) vacuum chamber, (k) TPC, (l) $^6$LiF beam catcher, and (m) turbo molecular pump.}
  \label{fig_bl05setup}
\end{figure*}

The experimental apparatus consists of two sections: 
the beam shaping section (b)-(e) and the detector section (f)-(m). 
In this experiment, the neutron beam is shaped to the spin flip chopper (SFC). Because the SFC requires polarized neutrons, the apparatus was installed downstream of Polarized-beam branch.
The SFC can create monochromatic bunches by combining the pulsed neutrons while avoiding $\gamma$-rays from upstream by shifting the beam axis. The SFC consists of magnetic super mirrors and neutron spin flippers~\cite{taketani2011high}, shown in (b)-(d) in Fig.~\ref{fig_bl05setup}. 

The neutron spin is controlled by switching RF current of the flippers. 
The spin flipped neutrons are passing through the magnetic mirrors and dumped, while the non-flipped ones are reflected by the mirrors to be transported downstream. 
The neutron beam is formed into bunches whose lengths are about half ($40\,$cm) of the TPC. 
Because the bunch intervals were set to $3.3\,$m to avoid overlapping of signal and background from the SFC or beam catcher, the number of bunches per pulse was adjusted to the allowable maximum of five. The contrast of the SFC ran into $\sim 400$~\cite{mishima2015neutron}.

Then neutron bunches are transported into the TPC (k) in Fig.~\ref{fig_bl05setup}, after passing through a beam monitor~\cite{ino2014} (e), a 50-$\rm \mu$m-thick Zr window (f), and the neutron switching shutter (g).
The shutter is a 5-mm-thick tile which is made of polytetrafluoroethylene (PTFE) containing 95\% isotopically enriched $^6$LiF with 30\,wt\%. The neutron transmission of the switching shutter is calculated as $3\,\times\,10^{-6}$.
The tiles are used to cover the inside of the beam duct (D) and the TPC, whose cross section is 40\,$\times$\,40\,mm$^{2}$ for the inlet of the TPC, and 60$\,\times\,$60 mm$^{2}$ for the outlet.
A very small part of the neutrons ($10^{-5}$--$10^{-6}$) make the neutron decays or {\Hecapture} in the TPC, and the rest of the beam is dumped at a beam catcher (l), which is a box filled with $^6$LiF powder with a 0.5\,mm PTFE window.

\subsection{Detector}\label{sec_detector}
\begin{figure}[ht]
	\begin{center}
	\includegraphics[width=0.7\columnwidth]{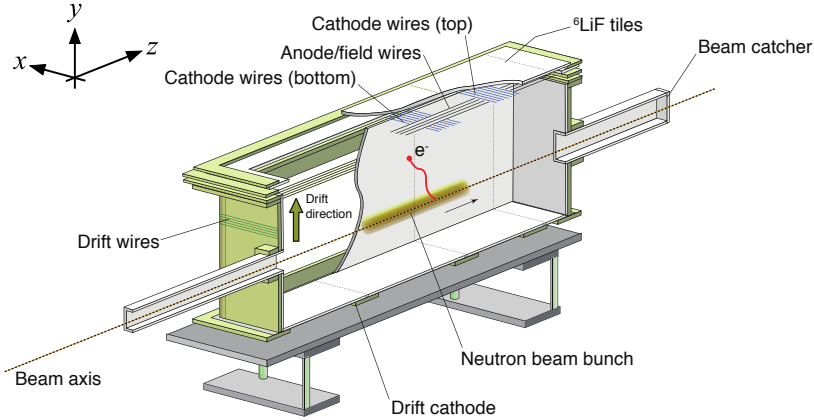}
	\end{center}
	\caption{Schematic view of the TPC~\cite{arimoto2015development}.}
	\label{fig_tpc3d}
\end{figure}

The TPC with polyether ether ketone (PEEK) and {\LiF}s was developed to detect neutron decays with a low background environment in the long-term operation~\cite{arimoto2015development}.
The schematic view of the TPC is shown in Fig.~\ref{fig_tpc3d}.
Since the count rate for the neutron decay is 1\,cps at 200\,kW in the beam bunches, 
that of the natural background ($S_{\rm const}$) should be kept at the same level or smaller for statistics. The PEEK frame is a substance with small radioactive material contamination. Thanks to this property, the background rate from the TPC support structure is suppressed to 4\,cps.

The inside of the TPC and the beam transport duct are covered with the {\LiF}s in order to avoid the background of $\gamma$-rays generated by neutrons hitting the wall. 
This {\LiF} can suppress the $\gamma$-ray generation against a neutron absorption to $2.3\,\times\,10^{-4}$~\cite{koga2020measurement}.
The {\LiF}s are packed in 100-$\rm \mu m$-thick PTFE sheets to prevent the ions emitted by the \Licapture\ reaction from entering the fiducial volume of the TPC. 
Almost all the scattered neutrons are absorbed by the {\LiF}s, therefore, possible $\beta$-nuclei produced in the TPC structure materials, which are the origins of $S_{\rm rad}$, are only $^8$Li (half-life 839.9\,ms, Q-value 16004\,keV) and $^{20}$F (half-life 11.07\,s, Q-value 7025\,keV)~\cite{firestone1996table}. Because the neutron absorption by the {\LiF} creates $^8$Li and $^{20}$F with probabilities of $2.5\,\times\,10^{-6}$ and $3.5\,\times\,10^{-5}$, respectively~\cite{arimoto2015development},
the difference of $S_{\rm rad}$ between $t$-foreground and background is estimated to be $2\,\times\,10^{-3}$.
These advantages enable us to achieve better statistical uncertainties than that of the previous measurement performed by Kossakowski $\it{et~al.}$ 

The TPC is installed in a vacuum chamber which is sealed with fluorocarbon O-rings.
A mixture of He and CO$_2$ of 85 and 15\,kPa as the TPC working gas was chosen because both of them have relatively small capture and scattering cross sections of the neutron. A few ppm of $^3$He is accurately admixed for the simultaneous measurement of the neutron flux. The working gas is used in the sealed condition during a series of measurements.

The TPC has a drift volume and a multi-wire proportional chamber (MWPC) placed above the drift volume. 
An aluminized PET film is placed on the {\LiF} at the bottom surface of the TPC and the drift voltage of $-9000\,{\rm V}$ is applied. On the surface of the {\LiF} at the top, additional aluminized PET films are placed and kept $+150$ and $+100\,{\rm V}$ to prevent the back-drifting of electrons outside the drift volume.
The MWPC consists of an anode plane sandwiched with cathode planes. The anode plane 
is made of anode and field wires which are stretched alternately in the $z$-direction 
with a spacing of 6\,mm. Each cathode plane has 162 wires stretched in the $x$-direction 
with a spacing of 6\,mm. The gaps between the anode and cathode planes are 6\,mm. 
The charge distribution of a particle track is projected onto the anode and cathode planes, 
and its two-dimensional image is obtained by measuring the signals from the anode/field wires and the cathode wires.
Table~\ref{tb:TPC} shows the specification of the TPC and each wire.
The details of the TPC are described in Ref.~\cite{arimoto2015development}.
\begin{table}[ht]
\centering
  \caption[Specification of the TPC]{Specification of the TPC and operating condition}
  \begin{tabular}{l|c}\hline
   Sensitive region & 290 mm ($x$) $\times$ 300 mm ($y$) $\times$ 960 mm ($z$) \\ 
   Anode         & 24 wires ($z$-direction), $\diameter$20\,$\mu$m AuW\\ 
   Field         & 24 wires ($z$-direction), $\diameter$50\,$\mu$m BeCu \\ 
   Cathode       & 162 wires\,$\times$2 ($x$-direction), $\diameter$50\,$\mu$m BeCu \\ 
   Gas mixture   & $^4$He : CO$_2$ : $^3$He $=$ 85\% : 15\% : 0.5--2 ppm\\ 
   Pressure      & 100 kPa\\ 
   Anode voltage & $+1720$ V\\ 
   Drift voltage & $-9000$ V\\ \hline
  \end{tabular}
  \label{tb:TPC}
\end{table}
A $^{55}$Fe X-ray source on a rotation stage is equipped at the side of the drift cage, and the $5.9 \,{\rm keV}$ X-rays are injected from two slits on the {\LiF} at 75 and 225\,mm from the MWPC for calibration of the TPC. 

The vacuum chamber is surrounded by a lead shield (i in Fig.~\ref{fig_bl05setup}) to reduce the environmental background radiation emitted from radioisotopes such as $^{40}$K, uranium-series, and thorium-series which are contained in the concrete of the building. 
The thickness of the lead is 5\,cm, which shields 98\% of environmental $\gamma$-rays.
Because $\gamma$-rays caused by neutron capture at the mirrors of the SFC produce considerable backgrounds, the shield thickness on the upstream side is 10\,cm.
Besides, 20-cm-thick iron walls (C in Fig.~\ref{fig_bl05setup}) are placed at the front and sides to shield $\gamma$-rays from the neighboring beamlines. 

A veto system using plastic scintillators (h in Fig.~\ref{fig_bl05setup}) is placed on the lead shield. It consists of 7 pairs of 12-mm-thick scintillator layers with wavelength-shifter fibers connected to 14 photomultiplier tubes.
The scintillators are arranged to surround all sides of the lead shields, except the bottom side. The coincidence of pairs of scintillators is used as a veto to cosmic-ray events. The veto efficiency is estimated to be 99\%. 
Finally, the whole count rate of $S_{\rm const}$ is suppressed to 8 cps without any cuts~\cite{arimoto2015development}. 

\begin{figure}[ht]
	\begin{center}
	\includegraphics[width=1.0\columnwidth]{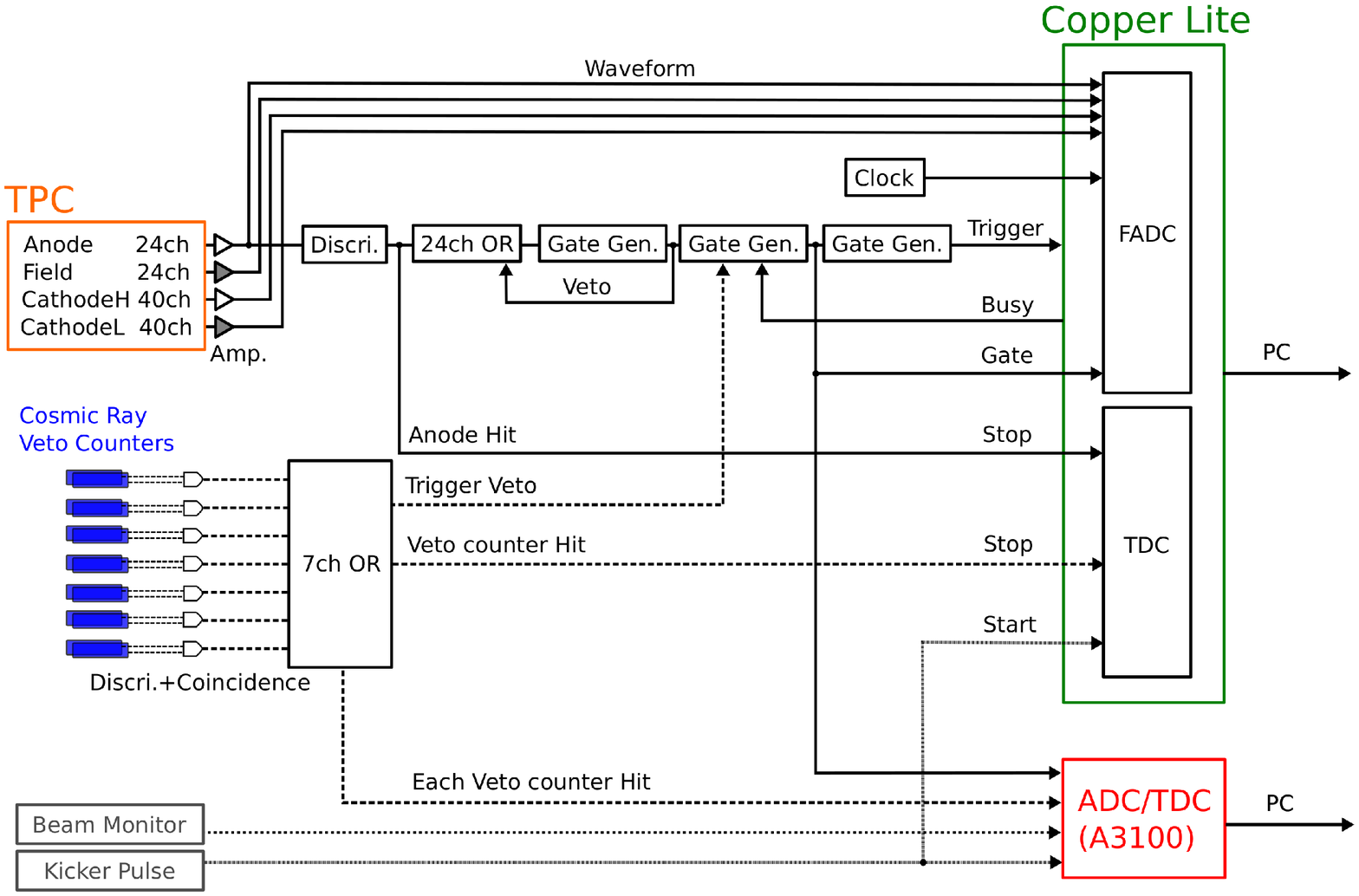}
	\end{center}
	\caption{Diagram of the DAQ.}
	\label{fig_DAQ}
\end{figure}

A diagram of the data acquisition system (DAQ) is given in Fig.~\ref{fig_DAQ}.
Signals of wires of the TPC are amplified and converted to voltages by preamplifiers.
The preamplifiers with two different gains are used to obtain a wide dynamic range; the anode and the bottom layer of the cathode wires with high gain, and the field and the top layer of the cathode wires with low gain.
The conversion factors of the high- and low-gain amplifiers are 1.3 and 0.23\,V/pC, respectively.
While each anode or field wire is connected to a readout channel, the four adjacent cathode wires are bundled into one readout channel.
A trigger for the DAQ is generated when at least one of the anode wire signals exceeds the threshold voltage of 20\,mV. The waveforms are recorded using a flash analog-to-digital converter (FADC) as data of 100\,$\rm \mu$s length with 100\,ns resolution. The waveforms of 70\,$\rm \mu$s after the trigger were treated as an event. Note that the number of triggers was recognized as the number of events. 
The measured time from the primary proton beam pulse (kicker pulse in Fig.~\ref{fig_DAQ}), which is referred to as $t$, is recorded by a time-to-digital converter (TDC).
The set of the FADC and TDC data is sent to a PC through the COPPER-Lite board, developed in KEK~\cite{igarashi2005common}. The information of the beam monitor, hit-timings of anode wires, cosmic-veto counters, and proton beam pulses is recorded in parallel by an ADC/TDC system (Nikiglass A3100). 

\subsection{Detector simulation}
A Monte Carlo code \verb|GEANT4| release 4.9.6.04~\cite{allison2016recent} is used for this experiment.
The physics lists of \verb|FTFP_BERT_PEN| and \verb|QGSP_BIC_HP| were employed to take into account the interaction of the low energy particles and the neutron capture reactions, respectively. 
The TPC, vacuum chamber, lead and iron shields, and cosmic-veto counters were included in the geometric condition of the simulation. 
The waveforms of the signals obtained from the anode, field, and cathode wires were simulated by calculating the drift motion of the ionized electrons which were liberated along the trajectories of the charged particles. 
Here, the number of ionized electrons was obtained from the local energy deposit and the $W$ value (40.9\,eV) for the gas mixture of 85\% He and 15\% CO$_2$. 
The non-linearity of the pulse heights due to the space charge effect in the electron avalanche process was taken into account using the saturation model~\cite{arimoto2015development,nagakura2018experimental}. 
The calculated event data were recorded and analyzed with the same procedure as the real experimental data.

\begin{figure}
	\begin{center}
	\includegraphics[width=0.6\columnwidth]{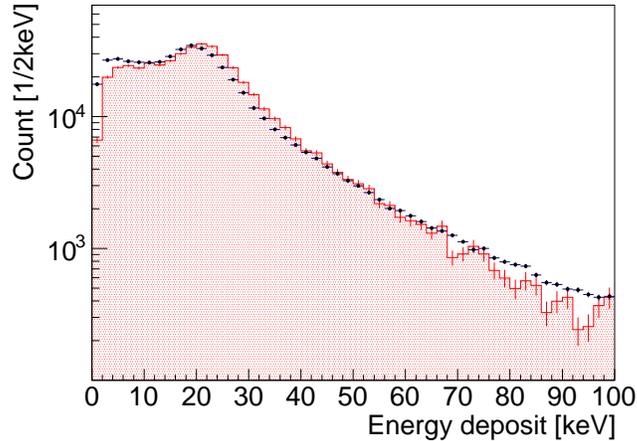}
	\caption{Energy deposit distribution of the cosmic-ray data (black dots) and simulation which was normalized by the total number of the measured histogram (red hatch).}
	\label{fig:energycomparison}
	\end{center}
\end{figure}

The conversion between the signal amplitude and the energy deposit was validated by comparing the measured and simulated spectra of cosmic muons.
The cosmic-ray veto signal from the coincidence of a pair of scintillators was occasionally inverted so that clear cosmic-ray events were acquired for monitoring the operating condition of the TPC by comparing the observed and simulated energy spectra of cosmic-rays as shown in Fig.~\ref{fig:energycomparison}.
The energy was calibrated by the $^{55}$Fe X-ray source, described in Sec.~\ref{sec_measurement}.
The discrepancy of the energy calibration in all of the measurement series was estimated to be 5--9\%, which is used to evaluate systematic uncertainties in cut energies of the event selection.

\subsection{Gas handling and $^3$He number density} \label{sec_gas}
\begin{figure}[ht]
\centering
 \hspace*{10mm}\includegraphics[width=0.7\columnwidth]{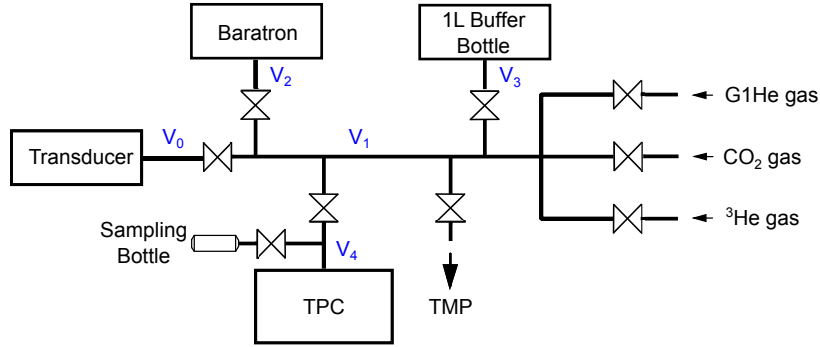}
  \caption{Schematic view of the gas handling system~\cite{HESJ_G3,kitahara2019_14N}.}
  \label{fig_gaspanel}
\end{figure}

Commercially available high purity He of 99.99995\% (G1He) and CO$_2$ of 99.999\% are used as the TPC working gas.
The neutron flux is measured by counting the {\Hecapture} reactions with $^3$He gas diluted in the working gas.
As shown in Eq.~(\ref{eq_tau_0}), since the measured neutron lifetime is a function of the number density of $^3$He, $\rho$, it should be determined with high accuracy.
The partial pressure of $^3$He was adjusted to 50--200\,mPa in order to obtain sufficient statistical accuracy in the neutron flux measurement through the detection of the {\Hecapture} reaction events. The maximum pressure of 200\,mPa was determined by the pileup of the {\Hecapture}, which was estimated as $\sim\,0.4\%$ with 200\,kW.
Because it is not easy to directly measure such a small pressure accurately, isotopically pure $^3$He gas ($>99.95\%$) was injected into a smaller container with high pressure ($\sim 3$\,kPa), and then released into the vacuum chamber of the TPC. The gas handling system for the procedure is shown in Fig.~\ref{fig_gaspanel}, where the details are described in 
Ref.~\cite{HESJ_G3,kitahara2019_14N}\footnote{These works have been done as an application of this experimental apparatus.}.
Here, the volume ratio of the vacuum chamber for the TPC to the small container was determined as $(1.497\,\pm\,0.028)\,\times\,10^{4}$ by measuring the pressure change when G1He gas was released from the container to the vacuum chamber. Corrections to the ideal gas law using the second virial coefficient and thermal transpiration effect on the transducer were taken into account.
The uncertainty of the ratio was evaluated based on the measurements of the pressure and the temperature, isotopic and chemical purity of $^3$He. 

\begin{figure}[ht]
	\begin{center}
	\includegraphics[width=0.7\columnwidth]{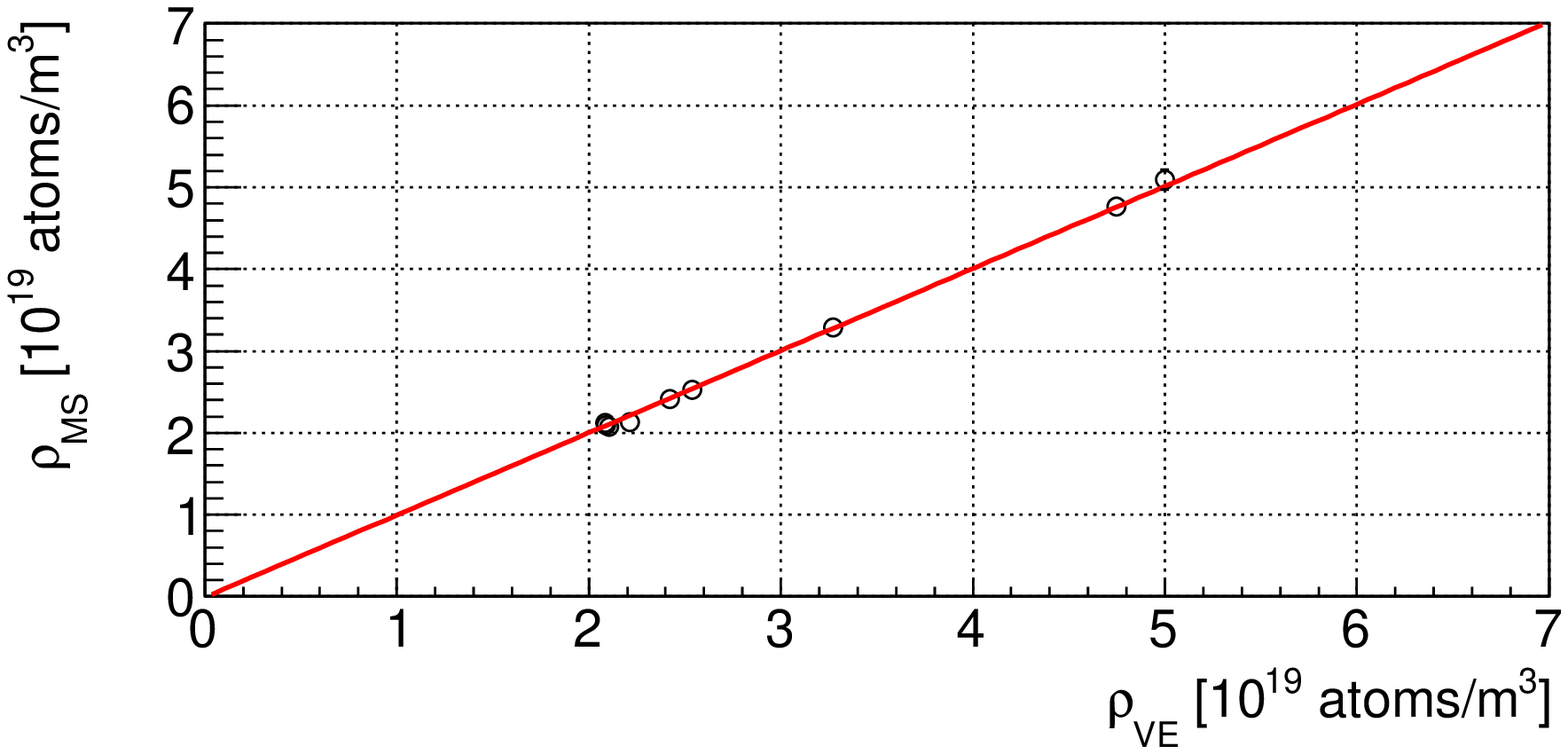}
	\end{center}
	\begin{center}
	\includegraphics[width=0.7\columnwidth]{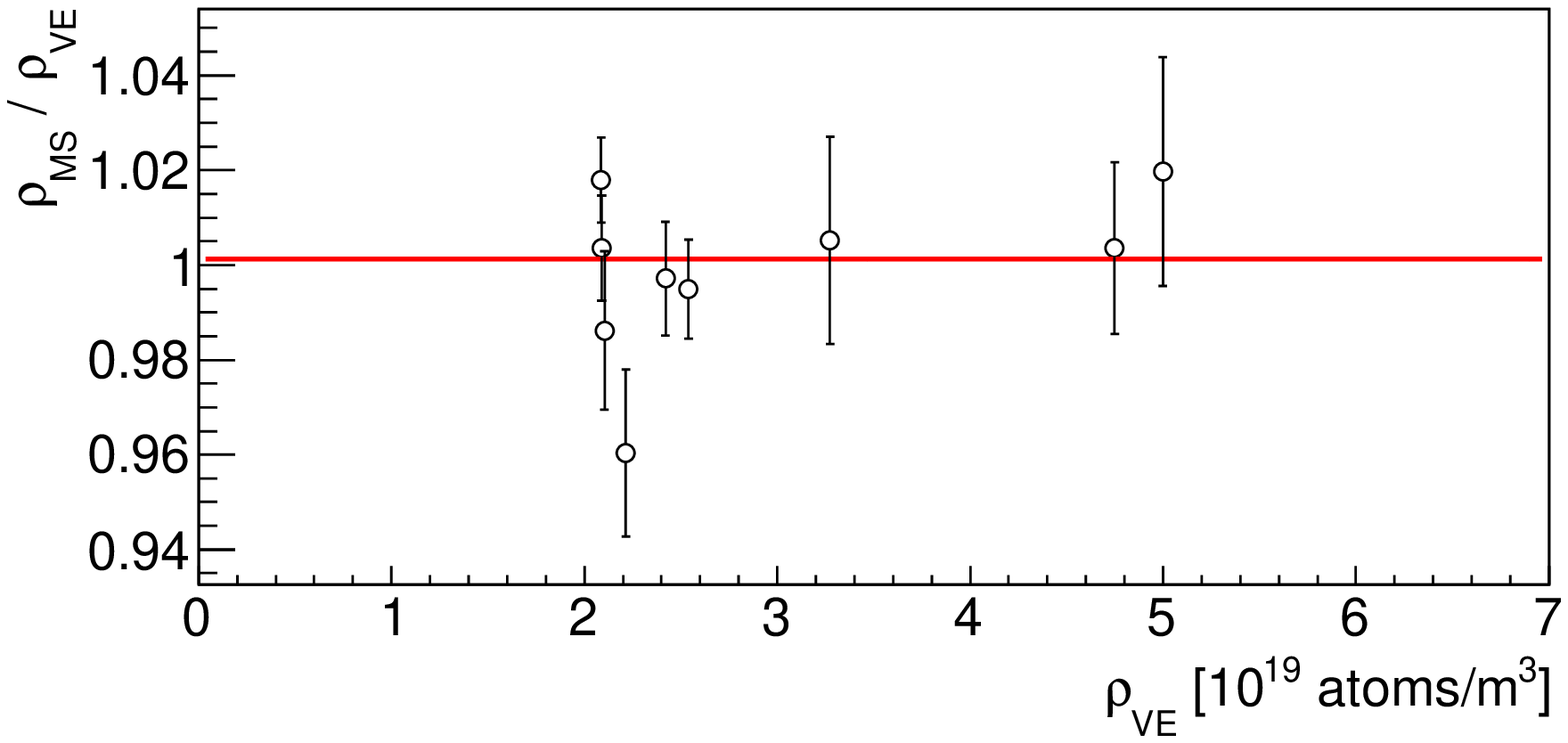}
	\end{center}
	\caption{
	The $^3$He number densities of $\rho_{\rm VE}$ on $x$-axis and $\rho_{\rm MS}$ on $y$-axis (top), and the ratio of the two methods (bottom).
	}
	\label{fig_MSdiff}
\end{figure}

Since $\rho$ in the working gas is a sum of the admixed $^3$He, $\rho_{\rm ad}$, and $^3$He in the G1He gas, $\rho_{\rm G1}$, we denote $\rho_{\rm VE}$ as 
\begin{equation}
\rho_{\rm VE} = \rho_{\rm ad} +  \rho_{\rm G1}.
\label{eq:rho}
\end{equation}
We determined $\rho_{\rm G1}$ by the ratio of $^3$He/$^4$He measured by a mass spectrometer~\cite{sumino2001highly} with accuracies of 1.5--3.0\% for all bottles used in this work. 
The working gases after the operation were sampled and their $^3$He/$^4$He ratios were measured by the mass spectrometer to confirm whether the $\rho$'s were properly controlled.
Putting the number density of $^3$He measured with the mass spectrometer as $\rho_{\rm MS}$, the relation between $\rho_{\rm VE}$ and $\rho_{\rm MS}$ is shown in Fig.~\ref{fig_MSdiff} for eight independent gas fillings. The values of $\rho_{\rm VE}$ and $\rho_{\rm MS}$ are consistent with the accuracy of $0.4\%$.
Because $\rho_{\rm VE}$ has better accuracy than $\rho_{\rm MS}$, we employ the $\rho_{\rm VE}$ as $\rho$.

The determined value of $\rho_{\rm VE}$ needs small corrections to be converted into $\rho$ during the operation. The vessel deformation due to the pressure and the temperature change was evaluated from the mechanical strengths and thermal expansion coefficients of structure materials of the chamber: stainless steel and aluminum.
We budgeted the correction as half of the maximum deformations with the symmetric uncertainty.
Another correction is for temperature non-uniformity.
A temperature gradient due to local heating around the preamplifiers at the top of the TPC was observed. 
It decreased the gas density of the high-temperature region and increased the others.
The increased amount of the $^3$He number density at the beam axis of the TPC was approximately 0.02\%~\cite{book2009heat}.
The number density $\rho$ was evaluated for each gas filling and applied for the analysis, and that of a typical gas filling is shown in Table~\ref{table_Helium3_NoA_Fill66}. 
As a result, the uncertainty of $\rho$ in the table was derived to be 0.42\%.

\begin{table}[htp]
\begin{center} 
\caption{
Value, correction, and uncertainty budgets of $\rho$ (Series 6)
}
\begin{tabular}{c c c c} \hline
Term & $^3$He number density    & Correction (\%)& Uncertainty (\%) \\
     & ($10^{16} \, {\rm m}^{-3}$) & \\\hline
$\rho_{\rm ad}$  & 2089 $\pm$ 7 &  & 0.3\\
$\rho_{\rm G1}$ &  202 $\pm$ 6  & & 3.0\\
$\rho_{\rm VE}$ & 2291 $\pm$ 9   &  & 0.4 \\\hline
Vessel Deformation (Pressure) &  & $-$0.15 & 0.15\\ 
Vessel Deformation (Temperature) &   & $-$0.02 & 0.02\\ 
Temperature uniformity &   &  & 0.02\\\hline\hline
$\rho$  &  2287 $\pm$  10 &  & 0.42\\\hline
\end{tabular}
\label{table_Helium3_NoA_Fill66}
\end{center}
\end{table}

\subsection{Measurement}\label{sec_measurement}
Six series of measurements were performed during the years of 2014 and 2016. At the beginning of every series, the TPC was refilled with fresh gas. In each series, the measurements with the beam shutter open and closed were repeated alternately. The period of each measurement was 1000 s. 
The total measurement times are summarized in Table~\ref{table_runs}. 
Note that $\rho$ values of Series 4 and 5 were intentionally changed to double and half, respectively, to check the systematic effect due to the $\rho$ values.

\begin{figure}
	\begin{center}
	\includegraphics[width=0.6\columnwidth]{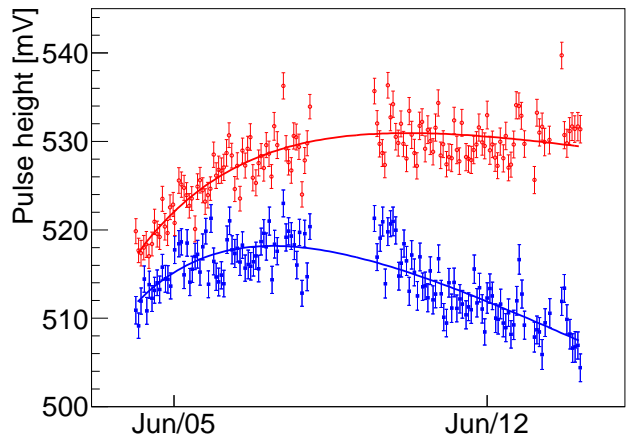}
	\caption{
	Pulse heights of the X-rays from the $^{55}$Fe source over time with fitting curves. The red circles and blue squares correspond to the source at 75 and 225\,mm from the MWPC, respectively.
	}
	\label{fig:energycalibration}
	\end{center}
\end{figure}

The fluctuation of the TPC gain was checked by the calibration runs with the $^{55}$Fe source placed at two positions on $y$-axis to measure attenuation in $y$-direction.
Figure~\ref{fig:energycalibration} shows the peak heights of the $5.9\,{\rm keV}$ X-rays as a function of the elapsed date from the beginning of a measurement series. It is expected that the fluctuation of the gain was caused by that of the temperature of the TPC. 
The drift velocity of the TPC was monitored by measuring the tracks of the cosmic-rays traversing from the top to the bottom of the TPC, which was stable at 0.3\% in a measurement series. The time differences of the earliest and latest signals in such events correspond to the maximum drift length, and the drift velocity averaged over the whole drift length was obtained as 1.0\,${\rm cm}/\mu$s with 4\% accuracy.
\begin{table}[thp]
\begin{center}
\caption{Summary of the measurement series}
\vspace{5pt}
\begin{tabular}{c c c c c} \hline 
Year  & Series &Beam power &   Measurement time  &  $^3$He number density ($\rho$)  \\ 
&        &(kW) & open/closed (hour) & ($10^{16}/{\rm m}^{3}$)\\\hline
2014  & 1      & 300  & 35 / 33     & 2417 $\pm$ 12 \\\hline 
2015  & 2      & 500  & 16 / 16     & 2084 $\pm$ 7 \\\hline 
      & 3      & 200  & 18 / 18     & 2348 $\pm$ 8 \\  
2016  & 4      & 200  & 73 / 69     & 4176 $\pm$ 13 \\ 
      & 5      & 200  & 69 / 63     & 1194 $\pm$ 8 \\ 
      & 6      & 200  & 71 / 71     & 2287 $\pm$ 10 \\\hline 
\end{tabular}
\label{table_runs}
\end{center}
\end{table}

\section{Analysis}\label{sec_ana}
\subsection{Procedure}
In this section, we describe the procedure to obtain the ratio of $S_{\beta}$, $S_{\rm He}$, $\varepsilon_{\beta}$, and $\varepsilon_{\rm He}$ in Eq.~(\ref{eq_tau_0}). 
The numbers of events, $S_{\beta}$ and $S_{\rm He}$, are derived from the experimental data, schematically shown in Fig.~\ref{fig:TOF}, by using the time-of-flight, open/closed of the neutron shutter, signal amplitude distribution, and track geometry together with the simulation of the detector response. 
The efficiencies, $\varepsilon_{\rm He}$ and $\varepsilon_{\beta}$, which are dependent on the cut conditions, are calculated by the simulation.

The neutrons arrived at the TPC generate the neutron decay and {\Hecapture} events. 
The CO$_2$ in the TPC working gas and nitrogen contamination in it cause {\Ccapture}, {\Ocapture}, and {\Ncapture} events, and, we denote them as $S_{\rm C}$, $S_{\rm O}$, and $S_{\rm N}$, respectively.
Neutrons scattered by the working gas or at the surface of the {\LiF} downstream the switching shutter additionally induce $\gamma$-rays by neutron captures of the structure materials.
We define the number of these events as $S_{\rm n \gamma}$.
These events appear accompanying the neutron bunches.
Finally, the number of the neutron-induced events in the TPC, $S_{\rm n}$, is given as
\begin{equation}
S_{\rm n}  = S_{\beta} + S_{\rm He} +  S_{\rm C} + S_{\rm O} + S_{\rm N} + S_{\rm n \gamma}.
\label{eq:Sn}
\end{equation}

The numbers of events observed in the foreground/background time region with the switching shutter open/closed are denoted as $S_{\text{FG-OPEN}}$, $S_{\text{FG-CLOSE}}$, $S_{\text{BG-OPEN}}$, and $S_{\text{BG-CLOSE}}$, respectively, which are normalized with the dead-time corrected time-windows and the incident neutron intensity measured with the beam monitor to match $S_{\text{FG-OPEN}}$.
The contents of $S$'s for these measurement modes are related to individual $S$ components via
\begin{eqnarray}
\scalebox{0.90}{$ 
\left[\begin{array}{c}
	S_{\text{FG-OPEN}} \\
	S_{\text{BG-OPEN}} \\
	S_{\text{FG-CLOSE}} \\
	S_{\text{BG-CLOSE}}
\end{array}\right]
= 
\left[\begin{array}{c}
	1 \\
	\eta_{\rm n}^{\rm SFC} \\
	\eta_{\rm n}^{\rm shutter} \\
	\eta_{\rm n}^{\rm SFC}\eta_{\rm n}^{\rm shutter}
\end{array}\right]
S_{\rm n}
+
\left[\begin{array}{c}
	1 \\
	\eta_{\gamma}^{\rm SFC} \\
	\eta_{\gamma}^{\rm shutter}  \\
	\eta_{\gamma}^{\rm SFC}\eta_{\gamma}^{\rm shutter}
\end{array}\right]
S_{\gamma}^{\rm mirror}
+
\left[\begin{array}{c}
    1 \\
    \eta_{\rm rad}^{\rm short} \\
    \eta_{\rm rad}^{\rm long} \\
    \eta_{\rm rad}^{\rm long}
\end{array}\right]
S_{\rm rad}
+
\left[\begin{array}{c} 
    1 \\
    1 \\
    1 \\ 
    1
\end{array}\right]
S_{\rm const},
$} 
\end{eqnarray}
where $\eta$'s are ratios for each component to $S_{\text{FG-OPEN}}$;
$\eta_{\rm n}^{\rm SFC}$ is the ratio of incident neutrons in the background to the foreground time region, and $\eta_{\gamma}^{\rm SFC}$ is the same one for neutron-induced $\gamma$-rays.
The ratios, $\eta_{\rm n}^{\rm shutter}$ and $\eta_{\gamma}^{\rm shutter}$, are the transmission of the switching shutter for the neutrons and $\gamma$-rays, respectively.
The ratios, $\eta_{\rm rad}^{\rm short}$ and $\eta_{\rm rad}^{\rm long}$, represent the residual radioactive isotopes, of which the background to foreground time region and shutter closed to open, respectively.

In this analysis, the following subtraction is performed to obtain $S_{\rm n}$;
\begin{eqnarray}
S_{\rm subt} &=& (S_{\text{FG-OPEN}} - S_{\text{BG-OPEN}}) - (S_{\text{FG-CLOSE}} - S_{\text{BG-CLOSE}})\nonumber\\
 &=& (1-\eta_{\rm n}^{\rm SFC} )(1-\eta_{\rm n}^{\rm shutter}) S_{\rm n}
+ (1-\eta_{\gamma}^{\rm SFC} )(1-\eta_{\gamma}^{\rm shutter}) S_{\gamma}^{\rm mirror}
+ (1-\eta_{\rm rad}^{\rm short}) S_{\rm rad}.
\label{eq:Ssubt}
\end{eqnarray}
Here, $\eta_{\rm n}^{\rm shutter}$ is negligibly small as described in Sec.~\ref{sec_beamtransport}, in contrast, $\eta_{\gamma}^{\rm shutter}$ is $\sim 0.95$. 
We can reasonably assume as 
$\eta_{\rm n}^{\rm SFC} \simeq \eta_{\gamma}^{\rm SFC}$, and 
$\eta_{\rm n}^{\rm SFC}$ is less than $5\,\times\,10^{-3}$ (see Fig.~\ref{fig:tEpm}), thus, we neglect them in this analysis.
According to the discussion in Sec.~\ref{sec_detector}, $(1 - \eta_{\rm rad}^{\rm short})$ is estimated to be $2\,\times\,10^{-3}$, then $(1 - \eta_{\rm rad}^{\rm short})S_{\rm rad}$ can be negligible because $S_{\rm rad}$ is $\sim 1/10$ of $S_{\beta}$ (see Fig.\ref{fig:tEpm} and later discussion). 
\begin{figure}
	\begin{center}
		\includegraphics[width=0.8\columnwidth]{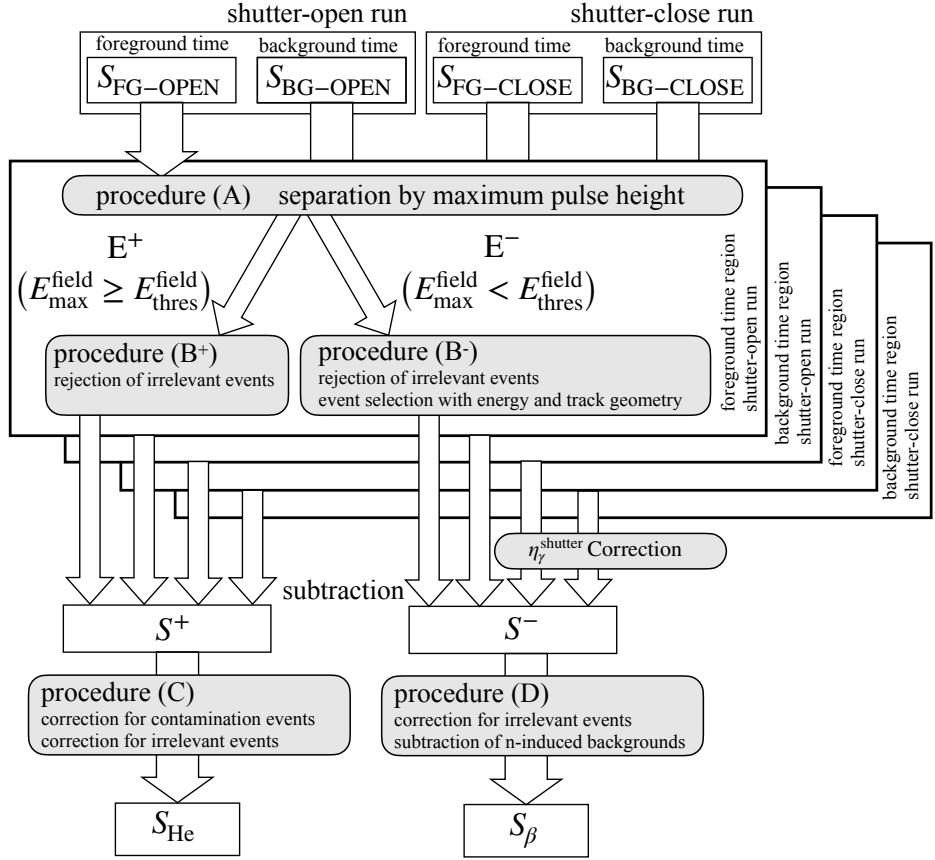}
	\end{center}
	\caption{
	Flowchart of the analysis procedure.
	Grey boxes stand for analysis procedures and white boxes for the event data.
	}
	\label{fig_eventselection_all}
\end{figure}
Consequently, Eq.~(\ref{eq:Ssubt}) can be written as
\begin{equation}
S_{\rm subt} \simeq 
S_{\rm n}  + (1-\eta_{\gamma}^{\rm shutter}) S_{\gamma}^{\rm mirror}.
\label{eq:Ssubt_simp}
\end{equation}
Here, the term with $\eta_{\gamma}^{\rm shutter}$
will be corrected by using simulations of the $\gamma$-rays from the neutron mirrors in the neutron transport in further analysis described in Sec.~\ref{sec:ProcB}.

A schematic diagram for the analysis procedures with cuts and corrections is shown in Fig.~\ref{fig_eventselection_all}. The procedures are as follows:
\begin{enumerate}
  \item First, the events are classified to high-energy group ($\rm E^{+}$) and low-energy group ($\rm E^{-}$) by using maximum pulse heights. The group $\rm E^{+}$ mainly consists of the {\Hecapture} events, and $\rm E^{-}$ contains the neutron decay events,
   described in procedure (A) in Sec.~\ref{sec:ProcA}.
  \item Individual cuts are applied to $\rm E^{+}$ and $\rm E^{-}$
  to extract the {\Hecapture} and the neutron decay events with higher purities,
  described as procedure (B$^{+}$) and (B$^{-}$) in Sec.~\ref{sec:ProcB}, respectively.
  \item The subtractions of ${\text{FG-OPEN}}$, ${\text {FG-CLOSE}}$, ${\text {BG-OPEN}}$, and ${\text{BG-CLOSE}}$ in Eq.~\ref{eq:Ssubt} are performed for $\rm E^{+}$ and $\rm E^{-}$ with the cuts to obtain $S^{+}$ and $S^{-}$.
  \item Corrections to exclude $S_{\rm N}$ and $S_{\rm O}$ are applied to $S^{+}$ in order to extract $S_{\rm He}$, described as procedure (C) in Sec.~\ref{sec:ProcC}.
  \item A correction to exclude $S_{\rm n \gamma}$ is applied to $S^{-}$ in order to extract $S_{\beta}$, described as procedure (D) in Sec.~\ref{sec:ProcD}.
\end{enumerate}
The detail of each procedure will be described below. 

\subsection{Region of the time-of-flight}\label{sec:tof_region}
\begin{figure}[ht]
	\begin{center}
		\hspace*{10mm}\includegraphics[width=0.7\columnwidth]{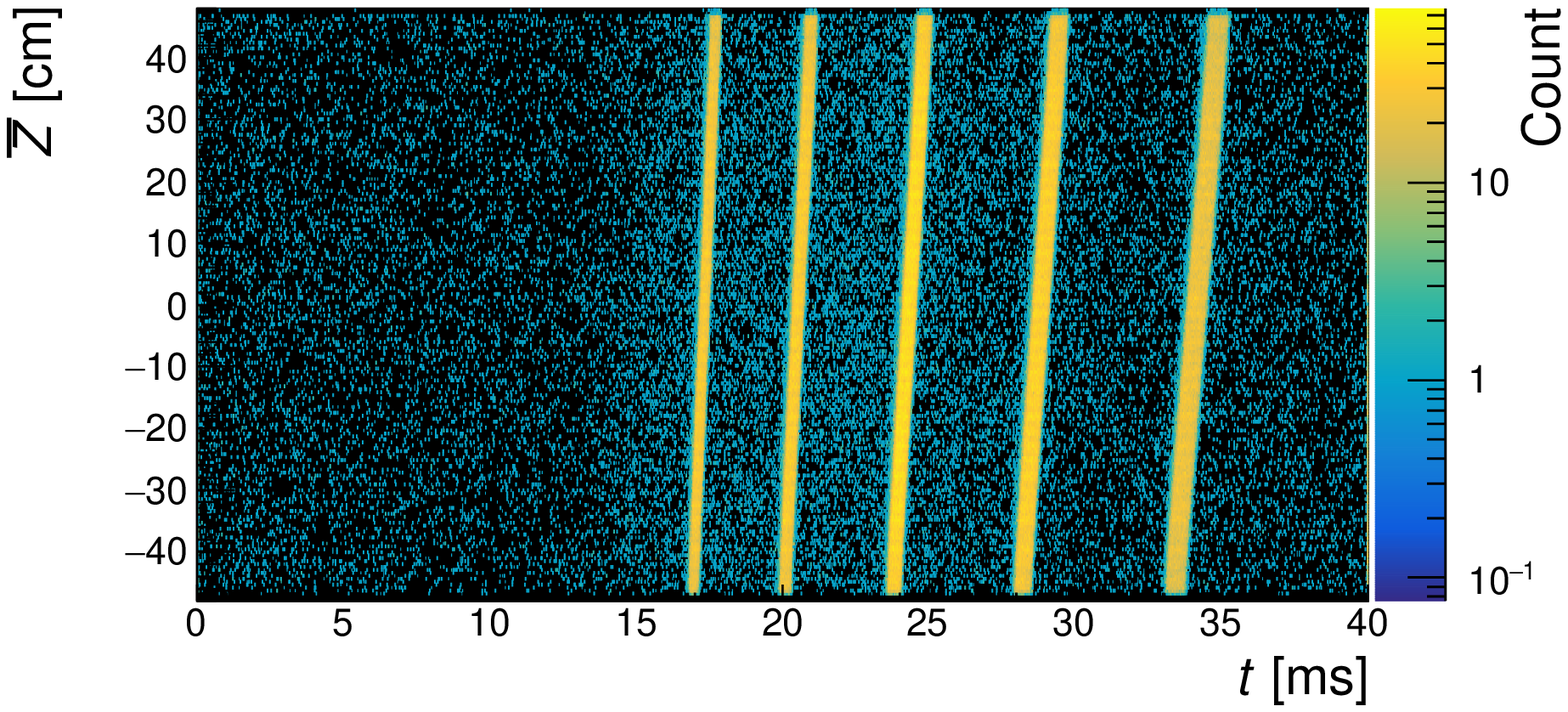}
       	\hspace*{10mm}\includegraphics[width=0.7\columnwidth]{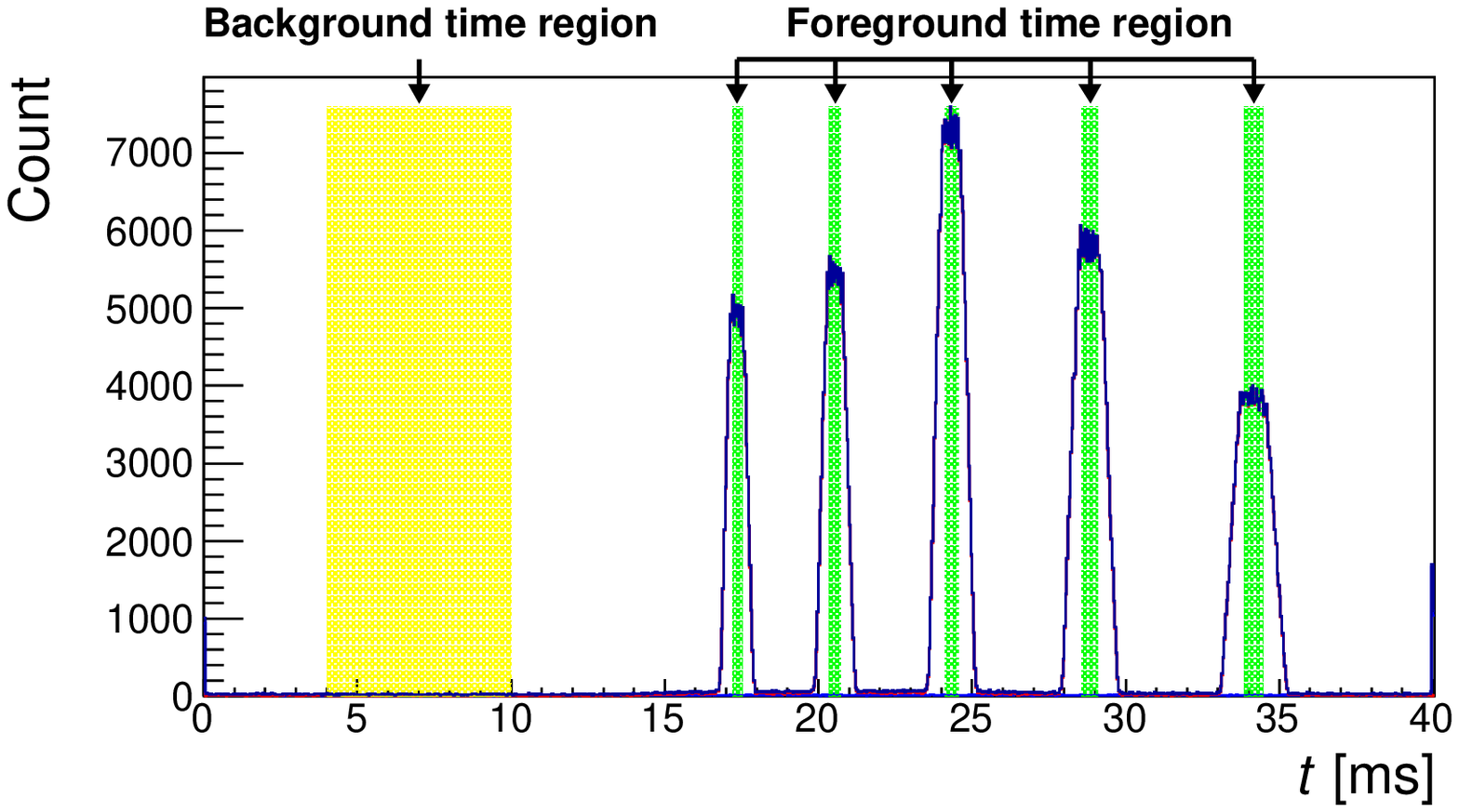}
	\end{center}
	\caption{
	Event distribution on the $\overline{Z}t$-plane (top) and its projection on to $t$-axis (bottom).
	The foreground and background time regions are hatched by green and yellow, respectively.
	}
	\label{fig_tof_raw}
\end{figure}

Since the tracks of the {\Hecapture} events are observed clearly in the TPC,
the event distribution observed with the low-gain amplifier outputs reflects the neutron distribution in the TPC.
Here, we define the weighted $z$-position as 
\begin{equation}
\overline{Z} = \frac
	{ \sum_{i} Q_{i} Z_{i} }
	{ \sum_{i} Q_{i} }
	,
\end{equation}
where $i$ is the channel number of a cathode wire, $Z_{i}$ is the $z$-coordinate of $i$-th cathode channel, and $Q_{i}$ is the charge on the low-gain amplifier of the $i$-th cathode channel.
Figure~\ref{fig_tof_raw} shows the distribution of experimentally observed events on the $\overline{Z}t$-plane.
The propagation of the five neutron bunches is clearly visualized as five bands. The slope on the $\overline{Z}t$-plane, which corresponds to the neutron velocity, decreased with $t$. Also, the time interval between bunches increased with $t$ since the bunches were made to be equally spaced as described in Sec.~\ref{sec_beamtransport}.

We defined the region of $-34\,{\rm cm} \le \overline{Z} \le 34\,{\rm cm}$ as the foreground region, which corresponded to the $t$-regions centered at 17.4, 20.5, 24.3, 28.8, and 34.2\,ms. The total foreground time width was 2.3\,ms. We defined the background region as $4\,{\rm ms} \le t \le 10\,{\rm ms}$ to minimize $S_{\gamma}^{\rm mirror}$.

\subsection{Procedure (A): Separation by maximum pulse height}
\label{sec:ProcA}
In further analysis, $S_{\rm subt}$ in Eq.~(\ref{eq:Ssubt_simp}) is divided into two groups; ion-emission events and the others, defined as $\rm E^{+}$ and $\rm E^{-}$ classes, respectively.
In the derivation process, some cuts are applied to remove the background and increase the purity of the signal, which are discussed in Sec.~\ref{sec:ProcB}.
Here,  
\begin{eqnarray}
S_{\rm subt}^\prime = S^{+} + S^{-}
\end{eqnarray}
is defined, where $S^{+}$ and $S^{-}$ are the numbers of events in $\rm E^{+}$ and $\rm E^{-}$ after the cuts, respectively.

\begin{figure}[ht]
	\begin{center}
	\includegraphics[width=0.6\columnwidth]{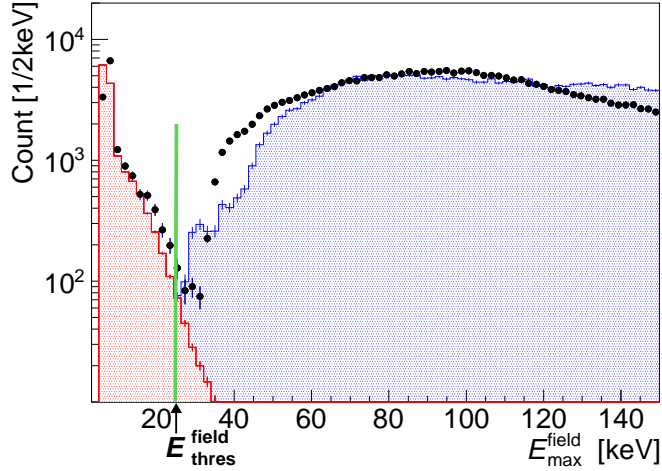}
	\end{center}
	\caption{
	Distribution of the maximum energy deposit among all field wires ($E_{\rm max}^{\rm field}$) with that of the simulation of the neutron decay (left red hatch) and the {\Hecapture} reaction events occurring (right blue hatch).
	The cut threshold (25\,keV) is also shown as a green vertical line.
	}
	\label{fig:hebetaseparation}
\end{figure}

Since the ion events have relatively higher energy deposits than electrons, each event is classified according to the maximum energy deposit among all the field wires, $E_{\rm max}^{\rm field}$.
Figure~\ref{fig:hebetaseparation} shows the $E_{\rm max}^{\rm field}$ distribution of $S_{\rm subt}^\prime$ together with the simulated distributions of $S_{\rm He}$ and $S_{\beta}$. The results of simulations show that the physical processes responsible for each event can be roughly classified and are mixed in the vicinity of their boundaries.
We set a threshold $E_{\rm thres}^{\rm field}=25\,{\rm keV}$ to minimize the admixtures between the two kinds of events as shown in Fig.~\ref{fig:hebetaseparation}. Because of the ambiguity of the {\Hecapture} simulation, the cut threshold was set lower than the valley of the measured spectrum. The events with $E_{\rm max}^{\rm field} \geq E_{\rm thres}^{\rm field}$ ($E_{\rm max}^{\rm field} < E_{\rm thres}^{\rm field}$) were classified as $\rm E^{+}$ ($\rm E^{-}$).

For the sake of simplicity, here we consider $S_{\rm He}$ and $S_{\beta}$ only,
and they are described as 
\begin{equation}
\left[\begin{array}{c}
	S^{+} \\
	S^{-}
\end{array}\right]
=
\left[\begin{array}{cc}
	1-\xi_{\rm sep}^{\rm He} &
	\xi_{\rm sep}^{\beta}
	\\
	\xi_{\rm sep}^{\rm He} &
	1-\xi_{\rm sep}^{\beta}
	\\
\end{array}\right]
\left[\begin{array}{c}
	S_{\rm He} \\ 
	S_{\beta} \\ 
\end{array}\right]
,
\label{eq:Smix}
\end{equation}
where $\xi$'s are the fraction of unfavored classification; 
$\xi_{\rm sep}^{\rm He}$ is the fraction of $S_{\rm He}$ mixed into ${\rm S^{-}}$
and $\xi_{\rm sep}^{\beta}$ is the fraction of $S_{\beta}$ mixed into ${\rm S^{+}}$. Note that the effects of them were less than 0.6\% for all measurements in this work.

\subsection{Procedure $({\rm B^{+}})$ and $({\rm B^{-}})$: Event selections for ${\rm E^{+}}$ and ${\rm E^{-}}$ classes} 
\label{sec:ProcB}
Respective cuts were applied to ${\rm E^{+}}$ and ${\rm E^{-}}$ classes as shown in Fig.~\ref{fig_eventselection_all}.
Thanks to the low radioactive TPC, the event rate caused by backgrounds other than the neutron bunch ones in $\rm E^{+}$ was suppressed to 0.15\,cps, which is $3\,\times\,10^{-3}$ of $S^{+}$. Therefore only a cut for electric noise was applied to ${\rm E^{+}}$, where the effect was negligibly small.

\begin{figure}
	\begin{center}
	\includegraphics[width=0.6\columnwidth]{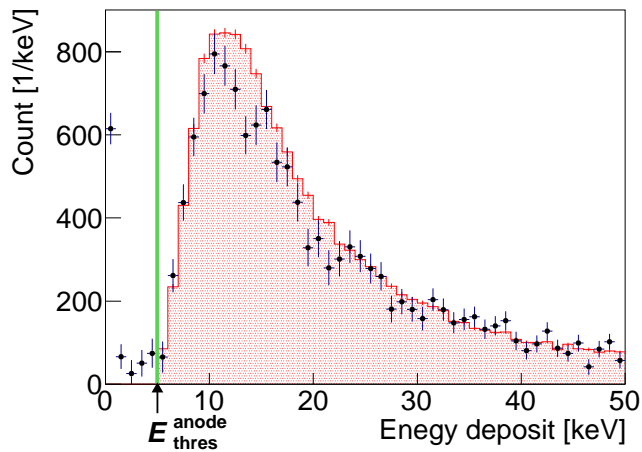}
	\caption{
	Energy distribution of $S^{-}$ of Series 6 (black circle) and that of the simulation of the neutron decay events normalized by the total events (red hatch). The green vertical line shows the cut threshold of 5\,keV.
	}
	\label{fig:AntiPoint}
	\end{center}
\end{figure}

Three cuts were applied to $\rm E^{-}$ described as follows.
The first cut is to remove the event by recoil nuclei from the {\Ccapture} reaction occurring in the TPC working gas which has the kinetic energy of 1.0\,keV. We set a cut on the energy deposit with threshold level $E_{\rm thres}^{\rm anode} = 5\,{\rm keV}$ to eliminate $S_{\rm C}$ from $ \rm E^{-}$.
The distribution of the energy deposit on the anode wires for $\rm E^{-}$ is shown in Fig.~\ref{fig:AntiPoint} together with the simulated spectrum of the neutron decay. The ratio of the residual of $S_{\rm C}$ after the energy cut to $S^{-}$, denoted as $\xi_{\rm C}$, was estimated to be less than $0.3\%$ by the Monte Carlo simulation.

The other two cuts were applied for statistical advantage by reducing $S_{\rm const}$ and $S_{\rm rad}$ components. 
Since the neutron decay events occur in the neutron beam region which is the center of the TPC, their spatial distribution in the TPC is different from that of the background events.
Thus, we can select the neutron decay events among various spatially distributed tracks using the waveform and/or distribution over the anode wires.
For $y$-direction, we required that the drift length is less than 190\,mm, which corresponds to the sum of the half-length of the TPC and the beam size, for removing charged particles generated outside of the beam region. For another background, $\beta$-decays of the tritiums (half-life 12.33\,years, Q-value 18.6\,keV)~\cite{firestone1996table} were observed, which had been produced by the \Licapture\ reactions in the TPC, and accumulated after a gas filling.
Since those decay electrons have short tracks and low energies, they have peaky shapes in their waveforms. Therefore, they can be identified by taking the ratio of the energy deposit around the highest peak to the full integration of the waveform. Events which had 80\% of the energy deposit in the peak region were rejected.

\begin{figure}[htbp]
	\begin{minipage}{0.48\hsize}
		\begin{center}
		\includegraphics[width=1.0\columnwidth]{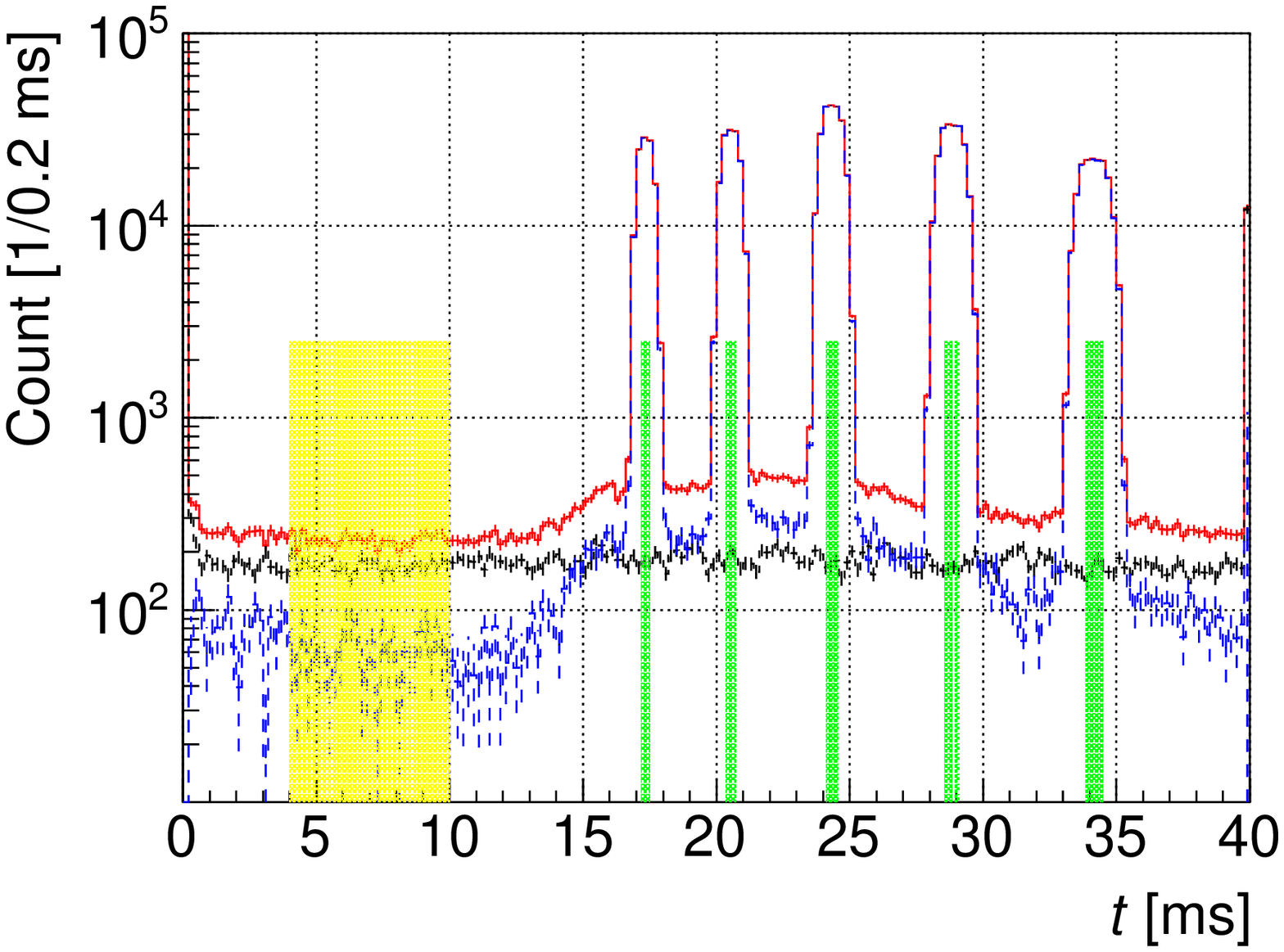}
		\end{center}
	\end{minipage}
	\begin{minipage}{0.48\hsize}
		\begin{center}
		\includegraphics[width=1.0\columnwidth]{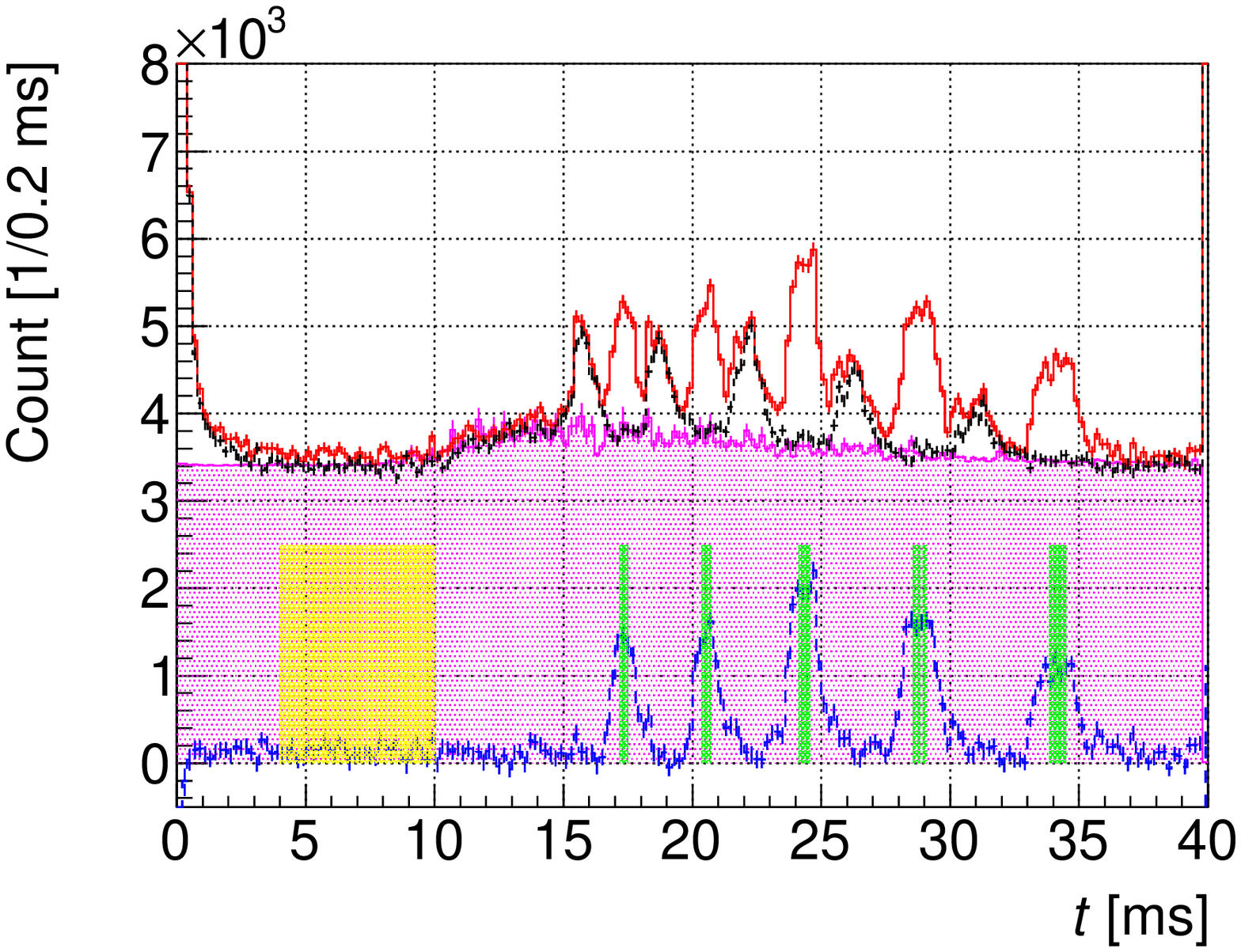}
		\end{center}
	\end{minipage}
	\caption{
	Time-of-flight spectra of the experimental data for $S^{+}$ (left) and $S^{-}$ (right).
	The red-solid and black-dotted lines represent the shutter open and closed data, respectively, and the blue-dashed one shows the difference between them. The hatched regions by green and yellow show the foreground and background time regions, respectively. The pink-hatched histogram is $S_{\gamma}^{\rm mirror}$ calculated by the simulations.
	}
	\label{fig:tEpm}
\end{figure}

The $t$-spectra of $S^{+}$ and $S^{-}$ after applying the cuts are shown in Fig.\ref{fig:tEpm}.
A simulation spectrum is plotted together with $S^{-}$.
In the simulation, the $\gamma$-rays produced from the neutron mirrors in the beam transport were calculated by \verb|PHITS|\,2.88~\cite{sato2018features} and the interactions of the $\gamma$-rays were simulated by \verb|GEANT4|.
The time-independent component was added to match the simulated $\gamma$-ray and  BG-CLOSED. 
The shielding effect of $\gamma$-rays by the neutron shutter, $(1-\eta_{\gamma}^{\rm shutter}) S_{\gamma}^{\rm mirror}$ in Eq.~(\ref{eq:Ssubt_simp}), was compensated here by using the simulation.
The difference between experimental data of FG-CLOSE and the simulation was budgeted as the uncertainty of the correction. The correction, denoted as $\xi_{\gamma}^{\rm shutter}$, was calculated to be ($0.3\,\pm\,0.3)\%$. 

\subsection{Procedure (C): Event selection and corrections for $S^{+}$}
\label{sec:ProcC}
\begin{figure}[htbp]
	\begin{minipage}{0.48\hsize}
		\begin{center}
		\includegraphics[width=1.0\columnwidth]{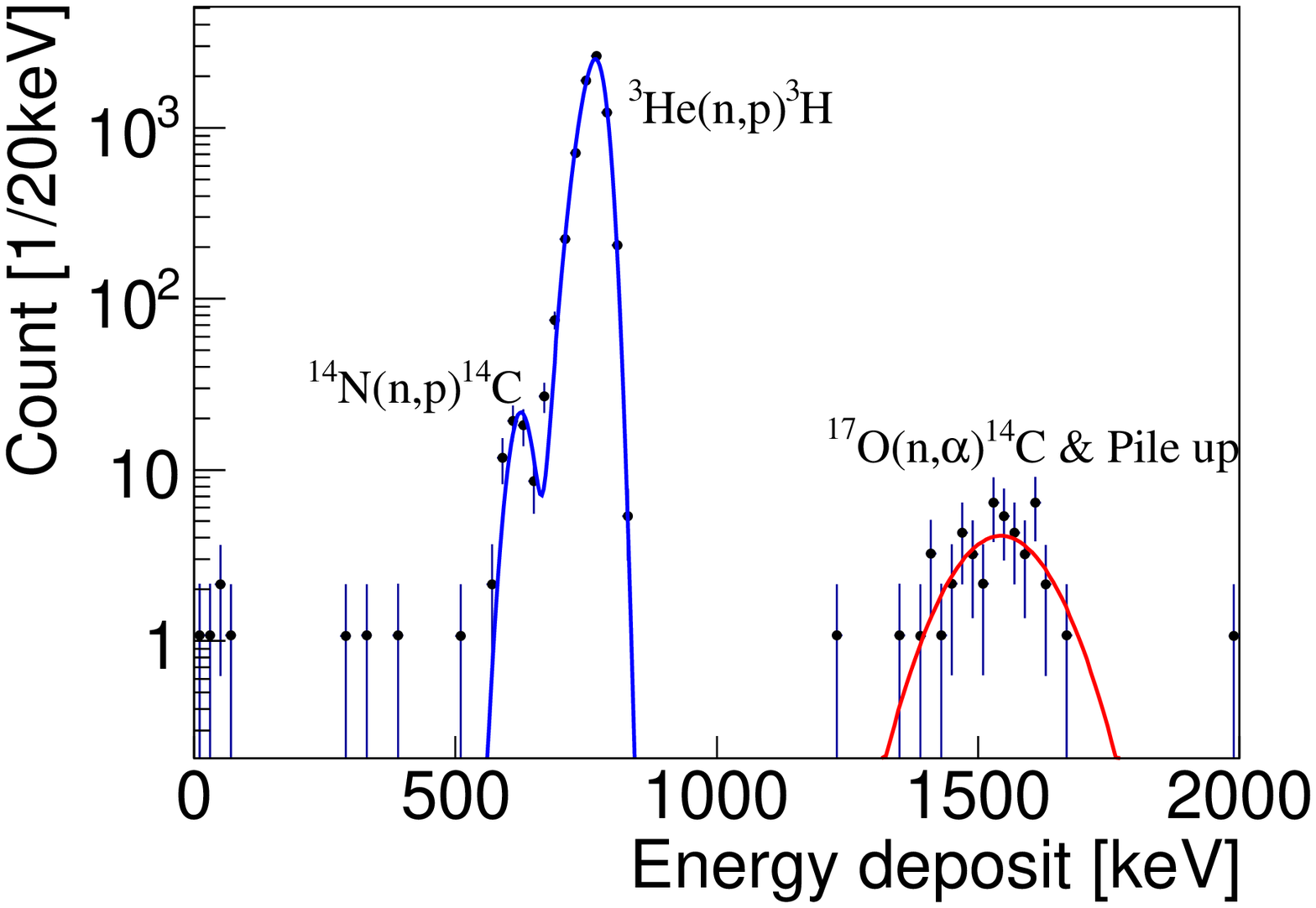}
		\end{center}
	\end{minipage}
	\begin{minipage}{0.48\hsize}
		\begin{center}
		\includegraphics[width=1.0\columnwidth]{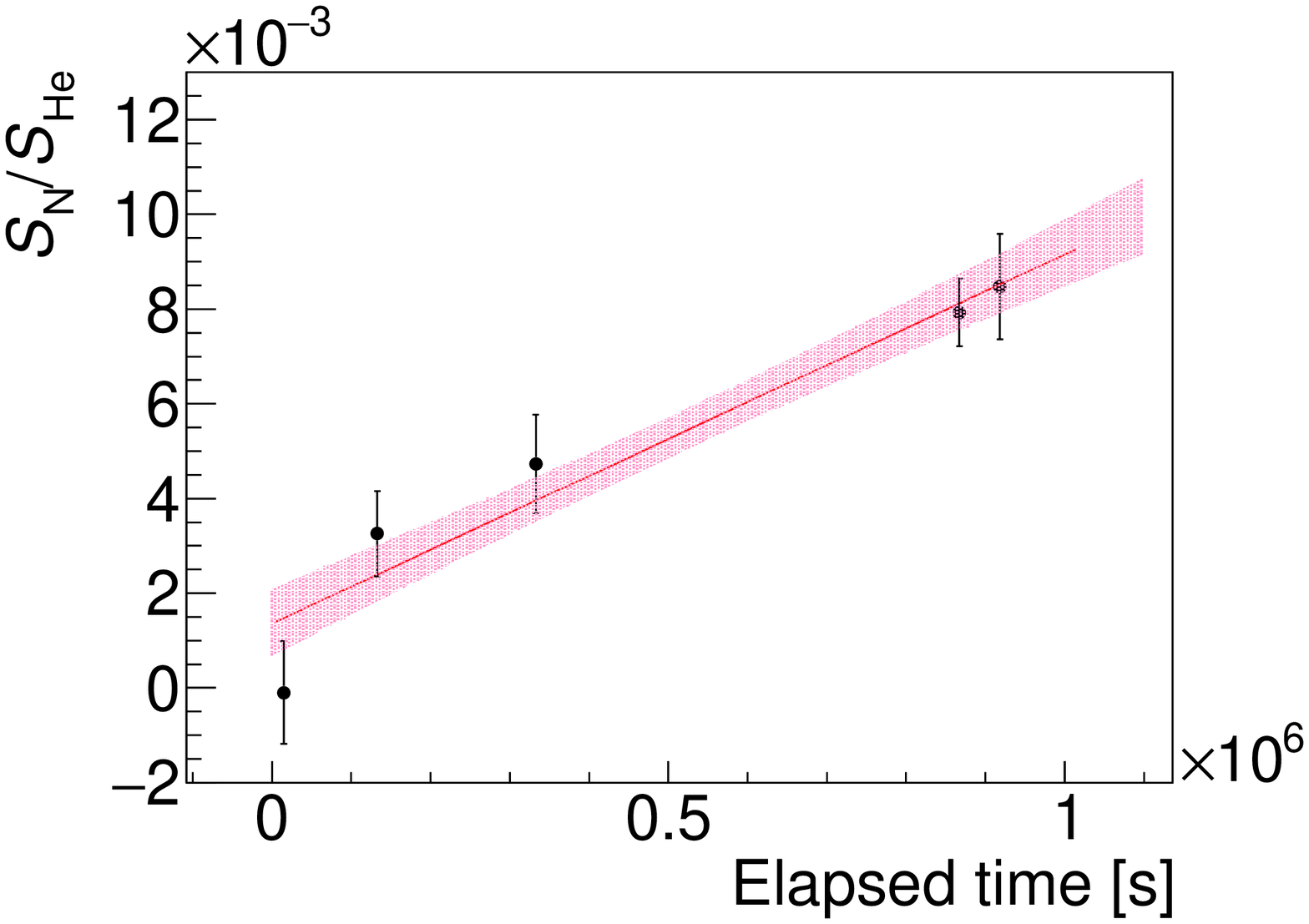}
		\end{center}
	\end{minipage}
	\caption{
	Pulse height spectrum of a low-gain operation (left) and ratio of event rates of the {\Ncapture} to those of the {\Hecapture} versus the elapsed time after gas filling (right). The red line and the hatched region show fitted curve and its error, respectively.
	}
	\label{fig:increaseN2}
\end{figure}

Contaminations of $S_{\rm N}$ and $S_{\rm O}$ are included in $S^{+}$.
Here, we define $S_{\rm Hecand}$ as
\begin{equation}
S_{\rm Hecand} =  S_{\rm He} + S_{\rm N} + S_{\rm O}  =  (1 + \xi_{\rm N} + \xi_{\rm O})S_{\rm He}
\end{equation}
with
$\xi_{\rm N} = S_{\rm N}/S_{\rm He}$ and
$\xi_{\rm O} = S_{\rm O}/S_{\rm He}$.
Because lowering the gain of the TPC was necessary to avoid the saturation of the pulse heights due to the space charge effect,
measurements with a reduced gain were performed every other day to monitor the influence of {\Ncapture}.
Figure \ref{fig:increaseN2} shows the pulse height spectrum of a low-gain operation and the ratio of the event rates of {\Ncapture} to {\Hecapture} as a function of the elapsed time. Because the outgas rate was roughly constant, the accumulated N$_{2}$ in the working gas proportionally increased over time.
Using the data for the time dependence of the {\Ncapture} event rate, $\xi_{\rm N}$ was estimated as (0.50\,$\pm$\,0.05)\%. This contamination level was consistent with a value expected from the N$_2$ concentration in the working gas which had been measured by gas chromatography. 

Since the {\Ocapture} reaction occurs with $^{17}$O nuclei contained in CO$_2$, which is the quenching gas of the TPC, its event rate can be estimated using the existing data of the isotopic abundance of $^{17}$O~\cite{berglund2011isotopic} and the {\Ocapture} reaction cross section~\cite{Mughabghab2006}.
The event rate ratio of {\Ocapture} to {\Hecapture} was evaluated as (0.51\,$\pm$\,0.03)$\%$. 

\begin{figure}
	\begin{center}
	\includegraphics[width=0.6\columnwidth]{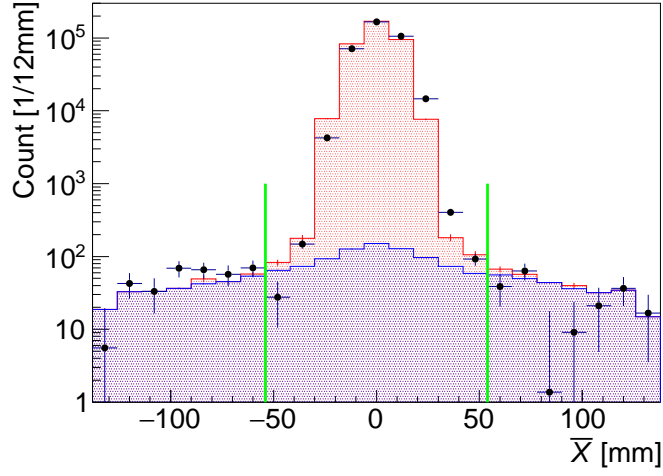}
	\end{center}
	\caption{
	Experimental data of $\overline{X}$ distribution (black dot) and simulations of incident neutrons (top red hatch) and scattered neutrons (bottom blue hatch).
    The green vertical lines show the boundaries of the incident and scattered neutrons.
	}
	\label{fig:fceResult}
\end{figure}

The incident neutrons were partially scattered ($\sim 1\%$) by the working gas or the entrance window of the vessel.
The scattered neutrons are captured by the {\LiF}s on the inner surface of the TPC or $^3$He, or decay in the path. 
Here, we define the average $x$-position of each event weighted by the energy deposit as
\begin{equation}
\overline{X} = \frac{\sum_{i} E_{i}^{\rm field} X_{i} } {\sum_{i} E_{i}^{\rm field}},
\label{eq_X_EC}
\end{equation}
where $X_{i}$ is $x$-position of $i$-th field wire with respect to the beam center.
The $\overline{X}$ distribution is shown in Fig.~\ref{fig:fceResult} and compared with the simulation of the {\Hecapture} events using the beam profile. 
The shape of the neutron beam was defined by the SFC geometry and collimators. The incident neutrons went into the beam catcher and were distributed in the $4\,{\rm cm}\,\times\,4\,{\rm cm}$ at $z=-34 \,{\rm cm}$ and $6\,{\rm cm}\,\times\,6\,{\rm cm}$ at $z=34\,{\rm cm}$.
The blue hatched area shows the simulation of the scattered neutrons, where the scattering distribution was calculated with the semi-classical model \cite{alcock1949neutron,alcock1951neutron}. 
Both simulations were scaled to the experimental data; the simulation for the scattered neutrons was normalized in the region of $|\overline{X}| > 54 \,{\rm mm}$, and the simulation for the incident neutrons was scaled so as to reproduce the experimental data together with the contribution of the scattered neutrons in the region of $|\overline{X}| \leq 54 \,{\rm mm}$. The ratio of the scattered neutrons to the incident neutrons, $\xi_{\rm scat}^{\rm He}$, was $0.39\,\pm\,0.04\%$, of which uncertainty is statistical error. We selected the events of the inside region of $|\overline{X}| \leq 54\,{\rm mm}$ for the further analysis.

A pileup event is defined as two or more events detected in the same time window (70\,$\mu$s). In this analysis, we classify an event to a type of event which caused the trigger, and the effect of the pileup is corrected so that the event number represents the trigger rate.
For the pileup correction of $S_{\rm He}$, there are three combinations of events; $\rm E^{+} \,\rm { to }\, \rm E^{+}$, $\rm E^{+} \,\rm { to }\, \rm E^{-}$, and $\rm E^{-} \,\rm { to }\, \rm E^{+}$. We classified these combinations as E$^{+}$ events because of their large energy deposit. The former two do not affect the result, because the event classification is correct. 
The last requires a small negative correction, $\xi_{\rm pileup}^{\rm He}$.
For conservative analysis, the magnitude and the uncertainty of it were set to be $0.08\%$ which corresponds to $\rm E^{-} \,\rm { to }\, \rm E^{+}$ pileup event rate.

Finally, $S_{\rm He}$ after the corrections described above is given as
\begin{eqnarray}
    S_{\rm He}
    &=&
    \frac{S_{\rm Hecand}}{(1+\xi_{\rm N}+\xi_{\rm O})}
     =  
    \frac{(1+\xi_{\rm pileup}^{\rm He})(S^{+} - \xi_{\rm sep}^{\beta} S_{\beta})}
         {(1+\xi_{\rm N}+\xi_{\rm O})(1+\xi_{\rm scat}^{\rm He})} \\\nonumber
    &\simeq&
    \left(
        1 - \xi_{\rm sep}^{\beta} \frac{S_{\beta}}{S^{+}} 
        -\xi_{\rm N}-\xi_{\rm O} -\xi_{\rm scat}^{\rm He} + \xi_{\rm pileup}^{\rm He}   
    \right) S^{+}.
\label{eq:SHe}
\end{eqnarray}
Corrections and uncertainties for $S_{\rm He}$ in Series 6 are summarized in Table~\ref{table_N3He_Fill66}. 
Note that $\xi_{\rm sep}^{\rm He}$ is not included because it is budgeted in $\varepsilon_{\rm He}$.
\begin{table}[htp]
\begin{center} 
\caption{Correction and uncertainty budgets of $S_{\rm He}$ (Series 6)}
\begin{tabular}{ c c c } \hline
Term &Correction (\%)& Uncertainty (\%) \\\hline
Statistics of $S^{+}$& & $\pm 0.18\,_{\rm stat}\,$\\\hline\hline
Misclassified neutron decay $(-\xi_{\rm sep}^{\beta} S_{\beta}/S^{+})$ & $-0.05$&$^{+0.05}_{-0.00}$ \\
Contamination of $^{14}$N $(-\xi_{N})$   &$-0.50$&$0.05$ \\
Contamination of $^{17}$O $(-\xi_{O})$   &$-0.51$&$0.03$ \\
Scattered neutron $(-\xi_{\rm scat}^{\rm He})$    &$-0.39$&$0.04$\\
Pileup ($\xi_{\rm pileup}^{\rm He}$)    &$-0.08$&$^{+0.08}_{-0.00}$\\\hline\hline
$S_{\rm He}$ & & $0.18\,_{\rm stat}\,^{+0.11}_{-0.06}\,_{\rm sys}$\\\hline
\end{tabular}
\label{table_N3He_Fill66}
\end{center} 
\end{table}

\subsection{Procedure (D): Background estimation and correction for $S^{-}$} \label{sec:ProcD}
\begin{figure}
	\begin{center}
	\includegraphics[width=0.5\columnwidth]{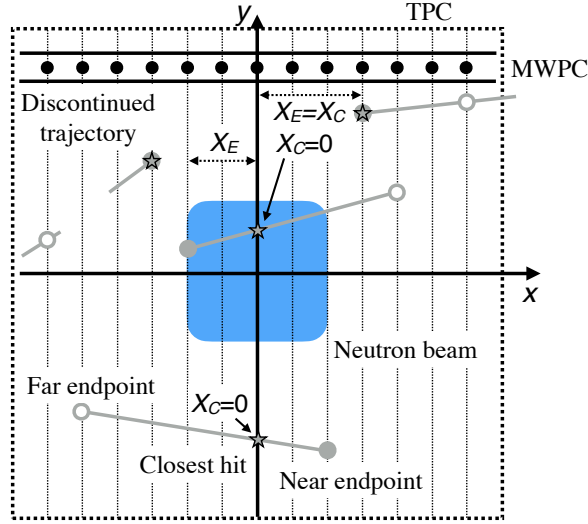}
	\caption{
	Schematic figure of tracks and anode hit positions to 
	illustrate the variables $X_{\rm C}$ and $X_{\rm E}$.
	The outermost dotted square region, inner blue-colored region,
	and upper black circles indicate the TPC, neutron beam region, and anode wires, respectively.
    The closed and open circles correspond to near and far endpoints from the central wire, and the star does the nearest hit position for each track.
	The number of wires and the geometric scale are not the same as those of the experiment.
	$X_{\rm C}$ is the distance along $x$-axis
	between the origin and the nearest hit anode wire,
	and	$X_{\rm E}$ is that between the origin and the nearer endpoint of the track.
	}
	\label{fig:Xdef}
	\end{center}
\end{figure}

\begin{figure}[htbp]
	\begin{minipage}{0.48\hsize}
		\begin{center}
		\includegraphics[width=1.0\columnwidth]{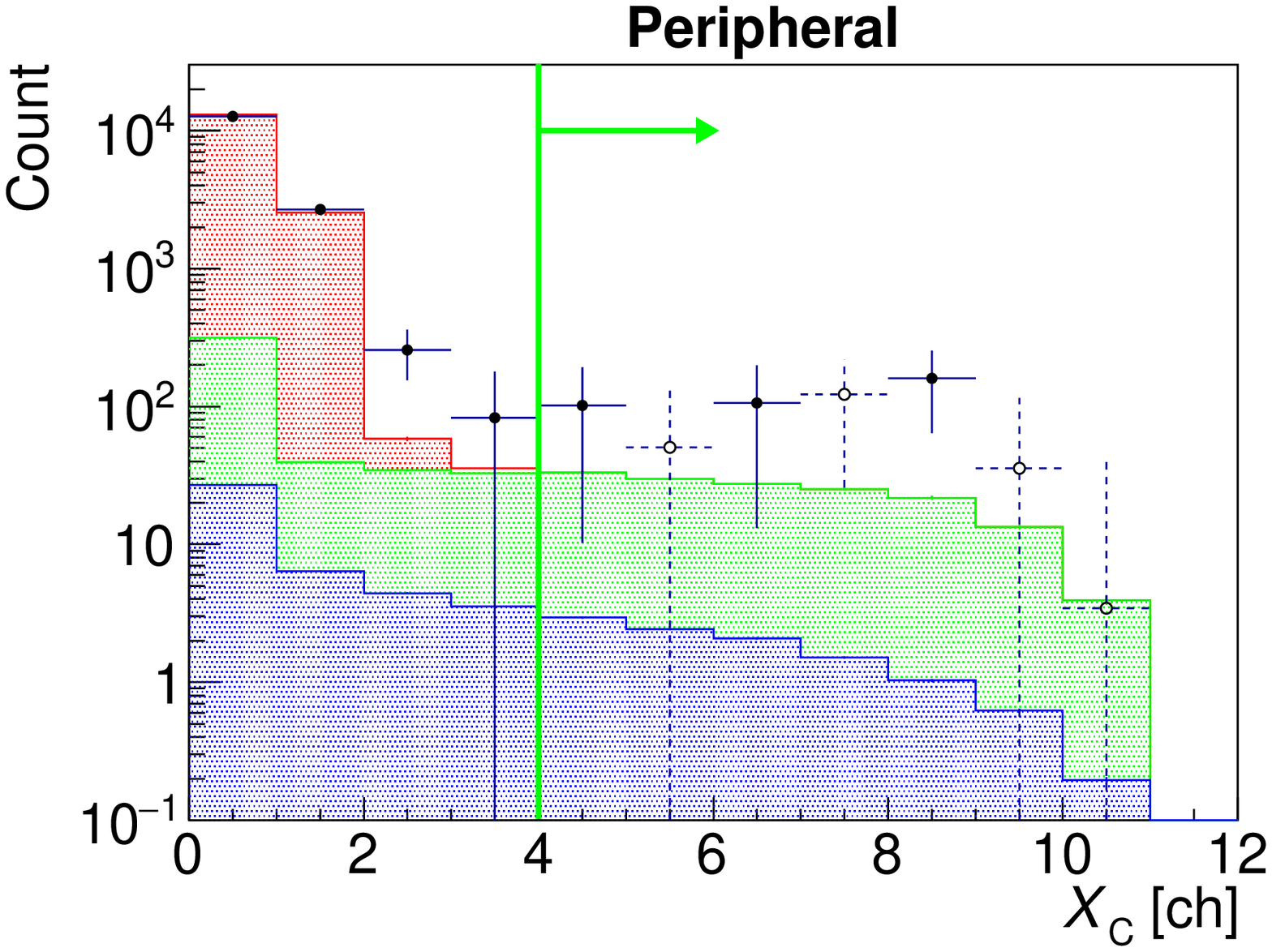}
		\end{center}
	\end{minipage}
	\begin{minipage}{0.48\hsize}
		\begin{center}
		\includegraphics[width=1.0\columnwidth]{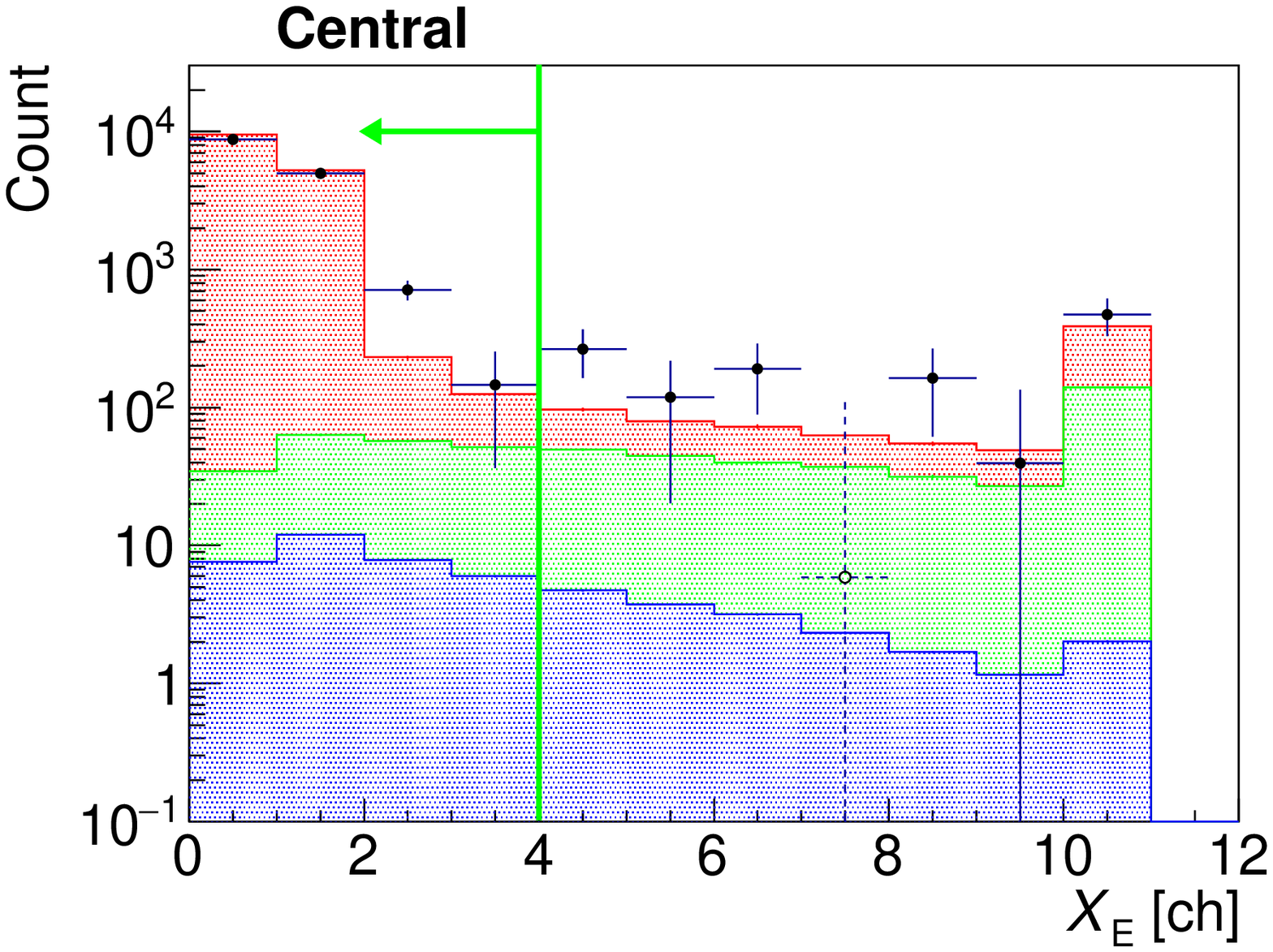}
		\end{center}
	\end{minipage}
	\caption{
	Distribution of $X_{\rm C}$ (left) and $X_{\rm E}$ (right) of $S^{-}$ (Series 6). Black circles with error bars show the experimental data, where the negative data points were turned up to show in the log plot as open circles with dotted error bars. The green vertical lines show the cut position and the green arrows indicate the central and peripheral components mentioned in the text.
	Simulated spectra of the neutron decays without scattering (red upper hatch), neutron-induced background (green middle one), and decay events of scattered neutrons (blue lower one) are plotted together. 
	}
	\label{fig:XCXE}
\end{figure}

Here, we define the event candidates of the neutron decay in $S^{-}$, $S_{\beta \rm cand}$ as 
\begin{equation}
S_{\beta \rm cand}  = S_{\beta} + S_{\beta \rm scat} + S_{\rm n\gamma},
\label{eq:Sbetacand}
\end{equation}
where $S_{\beta \rm scat}$ is the number of neutron decay events caused by the scattered neutrons, which can be estimated by $\xi_{\rm scat}^{\rm He}$ obtained in Sec.~\ref{sec:ProcC}. 
The neutron-induced $\gamma$-ray background, $S_{\rm n \gamma}$, is estimated by applying an analysis of track geometry and subtracted from the neutron decay candidate events.
Variables for the $x$-position of the anode wires, $X_{\rm C}$ and $X_{\rm E}$, are introduced for this analysis. A schematic figure for them is shown in Fig.~\ref{fig:Xdef}, where $X_{\rm C}$ is the distance along $x$-axis between the origin and the nearest hit anode wire, and $X_{\rm E}$ is that between the origin and the near endpoint of a track. The continuity of each track is not required in the analysis.
The distributions of $X_{\rm C}$ and $X_{\rm E}$ of $S^{-}$ are shown in Fig.~\ref{fig:XCXE} with scaled simulations of the neutron decay without scattering, the decay of the scattered neutron, and the neutron-induced $\gamma$-ray background.
Since the number of the anode wires is odd, the space for the 0-th channel is half of those for the other channels.

Using these variables, we classified the tracks as
the central $(X_\mathrm{E} \leq w)$,
the peripheral $(X_\mathrm{C} > w)$, and
the rest $(X_\mathrm{E} > w \,\mathrm{and}\, X_\mathrm{C} \leq w)$ components. 
The relation $X_\mathrm{C}\leq X_\mathrm{E}$ is always satisfied by definition.
Because tracks of the neutron decays in the beam have a hit within the neutron beam width, the neutron decay events without scattering are mainly classified in the central, and little ($<0.02\%$) exist in the peripheral. We can estimate $S_{\rm n\gamma}$ from the peripheral component with $S_{\beta \rm scat}$ determined by $\xi_{\rm scat}^{\rm He}$.
Ignoring the neutron decay without scattering in the peripheral, the central and peripheral components of $S_{\beta \rm cand}$ are  described as
\begin{eqnarray}
S_{\beta \rm cand}^{\rm cent}  &=& S_{\beta} + S_{\beta \rm scat}^{\rm cent} + S_{\rm n\gamma}^{\rm cent},
\\\nonumber
S_{\beta \rm cand}^{\rm per}  &=& S_{\beta \rm scat}^{\rm per} + S_{\rm n\gamma}^{\rm per},
\label{eq:Sbetacandcentper}
\end{eqnarray}
where 
$S_{\beta \rm scat}^{\rm cent}$, $S_{\beta \rm scat}^{\rm per}$, $S_{\rm n\gamma}^{\rm cent}$, and  $S_{\rm n\gamma}^{\rm per}$,
are the central and peripheral components of $S_{\beta \rm scat}$ and $S_{\rm n\gamma}$, respectively.
Note that $S_{\beta}$ represents the neutron decay without scattering. Though a small part of the $S_{\beta}$ were truncated by selections, the effects are compensated by the $\varepsilon_{\beta}$, discussed in Sec.~\ref{sec_Eff}. 
In this analysis, $S_{\rm n \gamma}^{\rm cent}$ is estimated by the simulation of $S_{\rm n\gamma}$ 
which is scaled so that $S_{\beta \rm scat}^{\rm per} + S_{\rm n\gamma}^{\rm per}$ matches the peripheral component of $S^{-}$.
Here, we define $\kappa$ as 
\begin{eqnarray}
S_{\rm n\gamma}^{\mathrm{cent}} = \kappa S_{\rm n\gamma}^{\mathrm{per}},
\end{eqnarray}
where $\kappa = 1.29$ by the simulation, 
then, $S_{\beta \rm {cand}}^{\rm cent}$ can be described as
\begin{eqnarray}
S_{\beta \rm {cand}}^{\rm cent} 
    = S_{\beta} + S_{\beta \rm scat}^{\rm cent} + \kappa S_{\rm n\gamma}^{\mathrm{per}}
    = S_{\beta} + (\xi_{\rm scat}^{\beta}+ \xi_{\rm n\gamma})S_{\beta \rm {cand}}^{\rm cent},
\end{eqnarray}
where $\xi_{\rm scat}^{\beta} = S_{\beta \rm scat}^{\rm cent}/S_{\beta \rm cand}^{\rm cent}$ and $\xi_{\rm n\gamma} = \kappa S_{\rm n\gamma}^{\mathrm{per}}/S_{\beta \rm {cand}}^{\rm cent}$.
The statistics of the peripheral component of $S^{-}$ and the systematics of $\kappa$ were budgeted as an uncertainty of $\xi_{\rm n\gamma}$.

As the average of all measurement series, $\xi_{\rm n\gamma}$ was $4.1\,\pm\,0.8\%$, which is 3.2-times of the expected value by the originally simulated (n,$\gamma$) reactions.
The origin of the difference is unknown but may be caused by
extra neutron captures outside the neutron shield of ${}^6\mathrm{LiF}$.
The unknown $\gamma$-rays which account
$(1-1/3.2)=0.69$
of $S_{\rm n\gamma}$ may obey different energy and position distributions from the simulation and result in a different $\kappa$.
Therefore, we used further track information in the peripheral component
to estimate the systematic deviation of $\kappa$.
In this estimation, we used the two sets of the simulations of $\gamma$-rays: 
one is the energy contrast distribution with the monochromatic energy 
$E_{\gamma}$ = 0.1, 0.2, 0.4, 0.8, 1.6,
$\cdots$, $12.8 \,\mathrm{MeV}$ 
and the same position distribution as the original simulation,
the other is the position contrast distribution which has the same energy distribution as the original simulation
and the initial position is the point where from the TPC center
moved on to the lead shield surface along one of the axis direction
$x\pm, y\pm, \,\text{or}\, z\pm$, which were selected to be the most biased position inside the shielding.

\begin{figure}[htb]
\begin{minipage}[t]{0.49\linewidth}
  \centering
  \includegraphics[width=\linewidth]{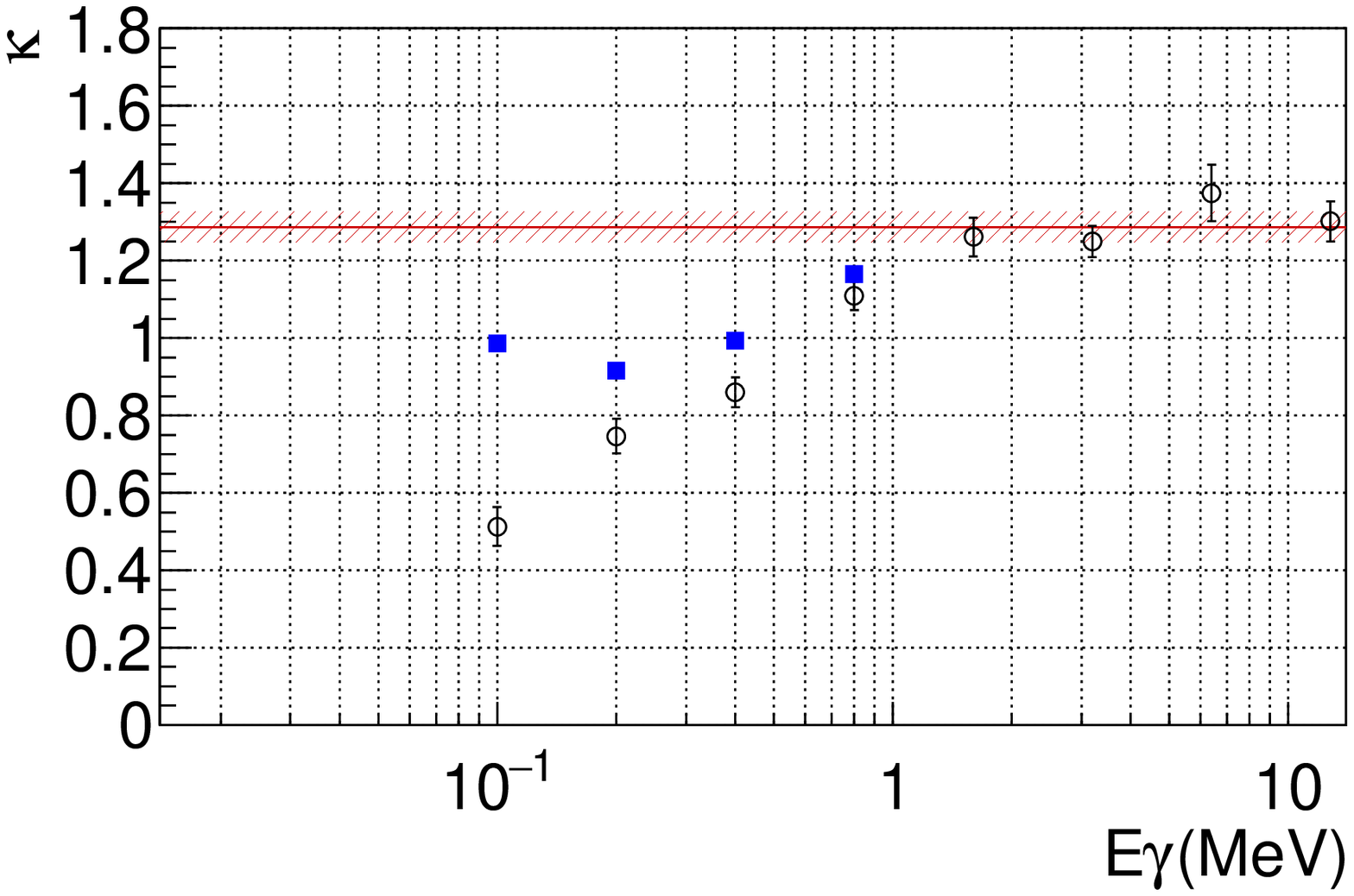}
\end{minipage}
\begin{minipage}[t]{0.49\linewidth}
  \centering
  \includegraphics[width=\linewidth]{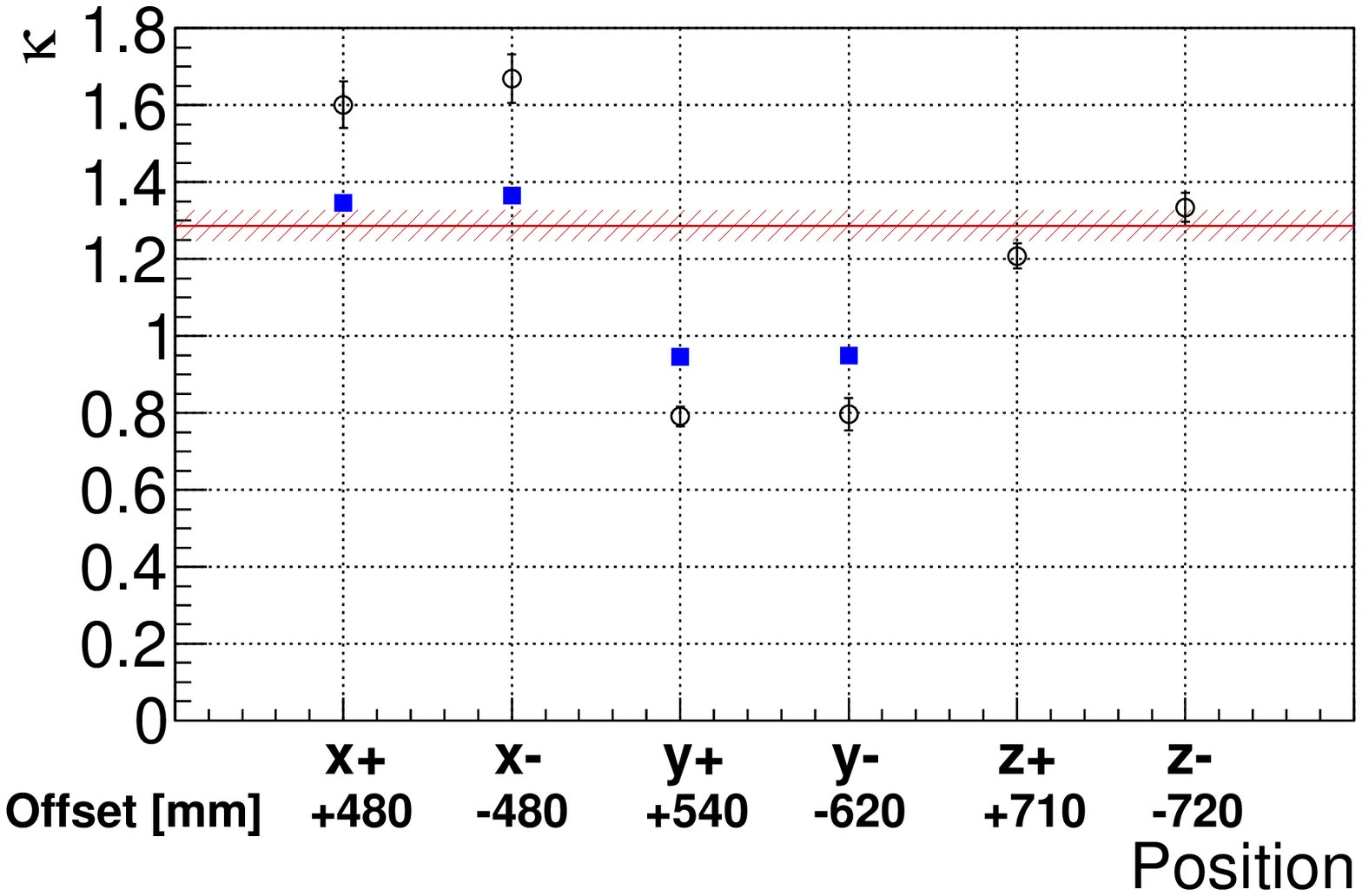}
\end{minipage}
  \centering
  \caption{Comparison of $\kappa$ of the original and energy contrast simulations (left) and position contrast ones with offset lengths of the source positions (right). The red line indicates $\kappa$ of the original simulation of $1.29\,\pm\,0.04\,_{\rm stat}\,$ and black circles does that of the contrast simulations. Blue squares correspond to the possible 1\,$\sigma$ deviations from the original $\kappa$ calculated by the minimum $\chi^2$ estimation. The resulting uncertainty on $\kappa$ is calculated using the maximum distance between blue square and red line.
  }
  \label{fig:kappa_ene_pos}
\end{figure}

\begin{figure}[htb]
\begin{minipage}[t]{0.49\linewidth}
  \centering
  \includegraphics[width=\linewidth]{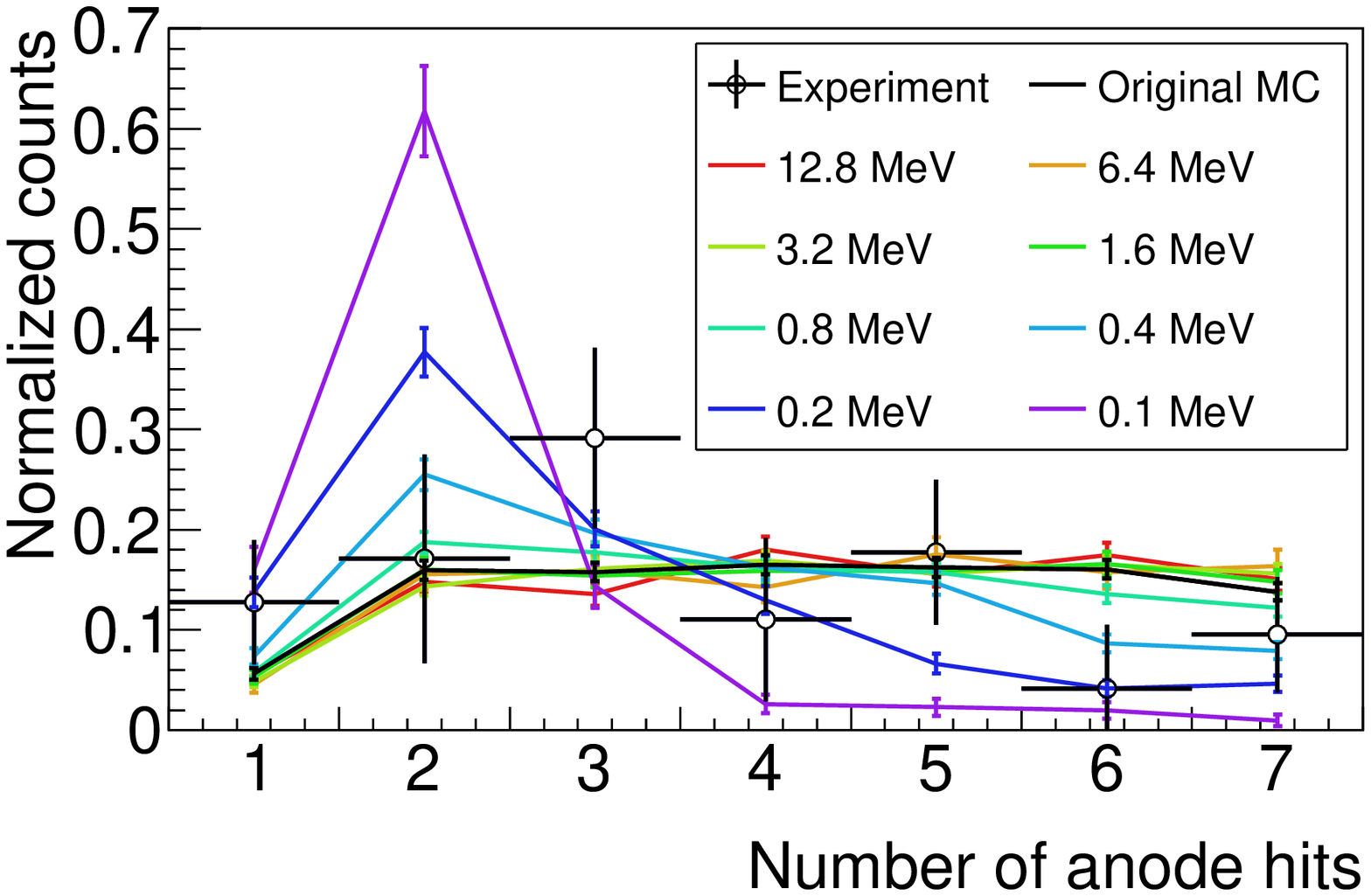}
\end{minipage}
\begin{minipage}[t]{0.49\linewidth}
  \centering
  \includegraphics[width=\linewidth]{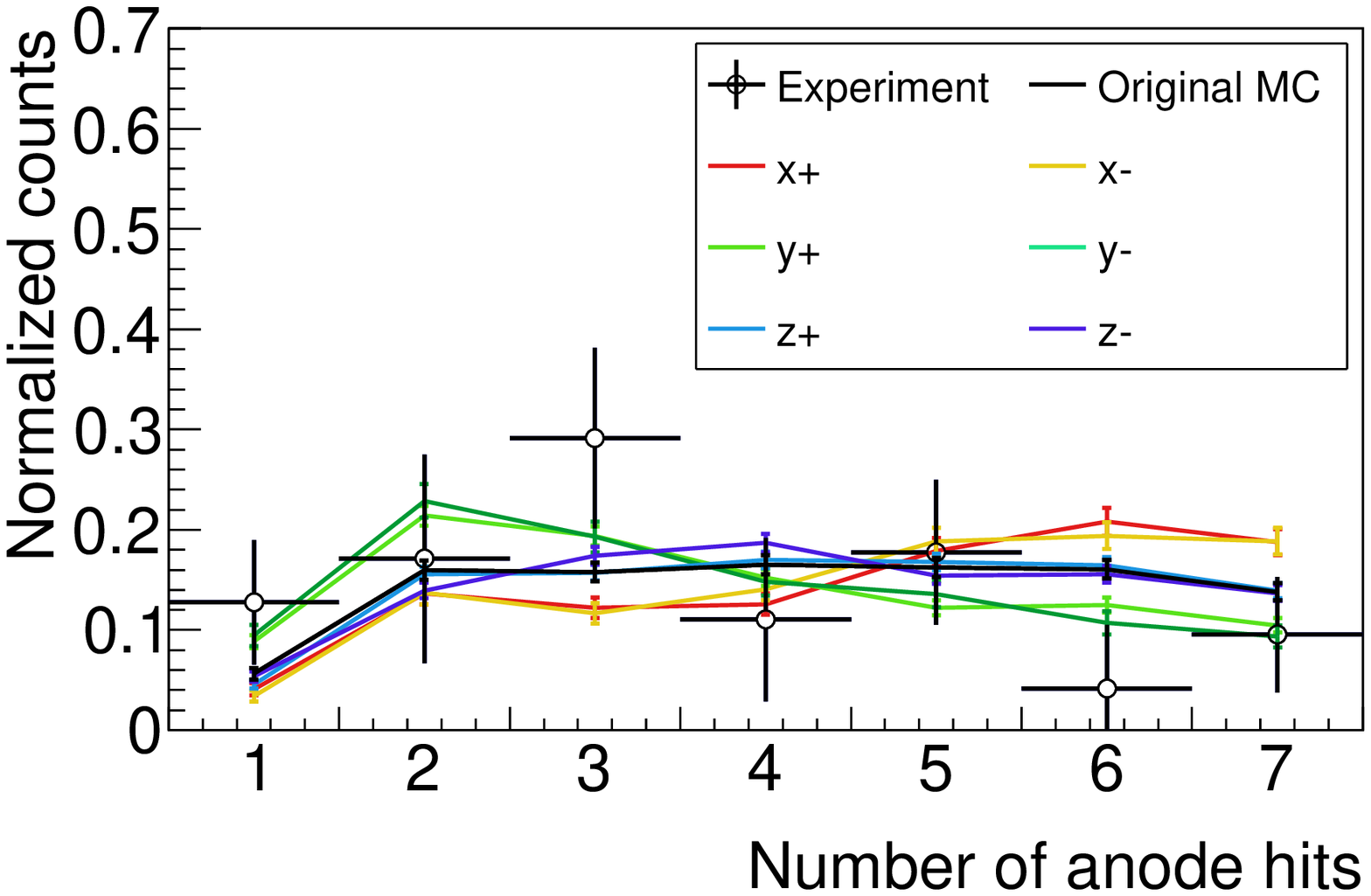}
\end{minipage}
  \centering
  \caption{Number of anode hits distributions of the experiment, the original, and energy contrast simulations (left) and position contrast ones (right) of the peripheral tracks. Black circles correspond to the distribution of the experiment. Black lines indicate that of the original simulation and other colored lines do that of contrast ones.
  }
  \label{fig:na_ene_pos}
\end{figure}

The $\kappa$ values calculated by the simulations shown in Fig.~\ref{fig:kappa_ene_pos}, and the anode wire hits distribution in the peripheral components shown in Fig.~\ref{fig:na_ene_pos}, are used to estimate the possible deviation of $\kappa$ from the original simulation. In the case of $E_{\gamma}$ = 0.1 MeV, $\kappa$ is $0.51\,\pm\,0.05$ but the spectrum of the anode distribution is unlikely from the experimental data as shown in the energy contrast simulation of Fig.~\ref{fig:na_ene_pos}. Hence, the contamination fraction of the $\gamma$-rays with that energy is constrained with the statistical range from the experimental data. The maximum possible value in 1$\,\sigma$ error of the contamination fraction, varied in the range of 0 to 0.69, was calculated by the minimum $\chi^2$ estimation. The results of the possible $\kappa$ values are shown in Fig.~\ref{fig:kappa_ene_pos} as the blue squares.
Since the energy contrast simulations of $E_\gamma \geq 1.6\,\mathrm{MeV}$ and the position ones of $z+$ and $z-$ have almost the same $\kappa$ as that of the original one,
we ignored them.
By taking the worst cases, the 1\,$\sigma$ deviation of $\kappa$ was obtained as 
\begin{eqnarray}
\kappa &=& 1.29\,\pm 
0.04\,_{\mathrm{stat}}\,
^{+0.00}_{-0.37}\,_{\mathrm{energy}}\,
^{+0.08}_{-0.34}\,_{\mathrm{position}}\\\nonumber 
&=& 1.29^{+0.09}_{-0.51}, 
\end{eqnarray}
where statistical and systematic errors were
summed in quadrature.

The pileup for $S_{\beta}$ was corrected in the same manner as described in Sec.~\ref{sec:ProcC}. If the neutron decay and the {\Hecapture} events were detected in the same time window, the events are possibly recognized as the {\Hecapture} events, and it reduces the number of neutron decay events. The probability calculated by the event rates was corrected. We also evaluated the pileups of events of $\rm E^{-} \,\rm { to }\, \rm E^{-}$, which might change the classification of the neutron decay events, or the other classes to the neutron decay events by changing its energy deposit and/or event topology. Thus, we budgeted the pileup probability of $\rm E^{-} \,\rm { to }\, \rm E^{-}$ as the systematic uncertainty. We denote this pileup correction as $\xi_{\rm pileup}^{\beta}$. 

Finally, $S_{\beta}$ after corrections described above is given as
\begin{eqnarray}
    S_{\beta}
    &=&
    (1 - \xi_{\rm scat}^{\beta} - \xi_{\rm n \gamma})S_{\beta \rm cand}^{\rm cent} \\\nonumber
    &=&
    (1 - \xi_{\rm scat}^{\beta} - \xi_{\rm n \gamma})
    \frac{(1+\xi_{\gamma}^{\rm shutter})(1+\xi_{\rm pileup}^{\beta})(S^{-} - \xi_{\rm sep}^{\rm He} S_{\rm He})}{(1+\xi_{\rm C})}
	\\\nonumber
    &\simeq&
    \left(
        1  - \xi_{\rm sep}^{\rm He} \frac{S_{\rm He}}{S^{-}} 
        -\xi_{\rm C} +\xi_{\gamma}^{\rm shutter} -\xi_{\rm scat}^{\beta} - \xi_{\rm n \gamma} + \xi_{\rm pileup}^{\beta}
    \right) S^{-}.
\label{eq:Sbeta}
\end{eqnarray}
Corrections and uncertainties for $S_{\beta}$ in Series 6 are summarized in Table~\ref{table_Nb_Fill66}.
Note that $\xi_{\rm sep}^{\beta}$ is budgeted in $\varepsilon_{\beta}$.
\begin{table}[htp]
\begin{center} 
\caption{Correction and uncertainty budgets of $S_{\beta}$ (Series 6)}
\begin{tabular}{c c c } \hline
Term &Correction(\%) & Uncertainty (\%) \\\hline
Statistic of $S^{-}$ & & 1.7$\,_{\rm stat}\,$\\\hline\hline
Misclassified ion events $(-\xi_{\rm sep}^{\rm He} S_{\rm Hecand}/S^{-})$ & 0.0 & $^{+0.0}_{-0.3}$ \\
Contamination of {\Ccapture} $(-\xi_{\rm C})$ & 0.0 & $^{+0.0}_{-0.3}$ \\
$\gamma$-ray shielding by neutron shutter $(\xi_{\gamma}^{\rm shutter})$ &$-$0.3 & 0.3\\
Scattered neutron ($-\xi_{\rm scat}^{\beta}$)    & $-$0.2 & 0.02\\
Neutron-induced $\gamma$-ray ($-\xi_{\rm n\gamma}$) &  $-$1.3  &2.0$\,_{\rm stat}\,^{+0.5}_{-0.1}\,_{\rm sys}\,$\\
Pileup $(\xi_{\rm pileup}^{\beta})$ & $+0.2$&$^{+0.4}_{-1.2}$ \\\hline\hline
$S_{\beta}$ &  &2.6$\,_{\rm stat}\,^{+0.6}_{-1.3}\,_{\rm sys}\,$\\\hline
\end{tabular}
\label{table_Nb_Fill66}
\end{center} 
\end{table}

\subsection{Efficiency $\varepsilon_{\rm He}$ and $\varepsilon_{\beta}$} \label{sec_Eff}
The detection efficiencies, $\varepsilon_{\rm He}$ and $\varepsilon_{\beta}$ in Eq.~(\ref{eq_tau_0}), were calculated by the simulation.
Since the trigger inefficiencies for the neutron decay and {\Hecapture} without scattering were estimated to be small enough from the simulation ($<10^{-3}$ and $<10^{-4}$, respectively), the systematic uncertainties of the efficiencies were evaluated for the event selections described in the former subsections.
We summarized the results of the cut efficiencies and uncertainties of $\varepsilon_{\rm He}$ and $\varepsilon_{\beta}$ in Table \ref{table_Eff3He_Fill66} and \ref{table_Effb_Fill66}. 
The value in the efficiency column for each cut means the ratio of the neutron decay which was rejected when only the corresponding cut was applied.
 
The uncertainties of the cut efficiencies were estimated by taking into account possible deviations of the cut thresholds. For instance, the discrepancy between the measurement and the simulation in the energy spectra of the cosmic-rays for all measurement series was 5--9\% as shown in Fig.~\ref{fig:energycomparison}. 
Hence we accounted for a change when the cut threshold in energy was shifted by the discrepant value as a cut uncertainty. The uncertainties for the $E_{\rm max}^{\rm field}$ cut and the low energy cut at $E_{\rm thres}^{\rm anode}$ were obtained in this way.
For the track geometry ($y$-direction), the non-uniformity of 9.5\% of the drift velocity was considered. The uncertainty for the tritium cut was estimated by the standard deviation of the peak/full-integration ratio distribution of the $5.9\,{\rm keV}$ X-ray waveforms.
The uncertainty of the threshold of $X_E$ could be caused by the misalignment of the beam position, which was estimated to be less than 2\,mm. We budgeted 12\,mm, which corresponds to 1 wire width, for the conservative estimation.

Some other systematic effects on $\varepsilon_{\beta}$ are discussed here. The electrons emitted from the neutron decay have an angle distribution around the neutron polarization. The angular distribution, $W(\theta)$, can be described as
\begin{eqnarray} \label{eq:PolarizedBetaDirection}
W(\theta) = 1 + \frac{v}{c}PA\cos(\theta), 
\end{eqnarray}
where $\theta$ is an angle between the direction of the electron and neutron polarization, 
$v$ is the velocity of the electron, 
$c$ is the speed of light, 
$P$ is the polarization of the neutron, and 
$A$ is the asymmetry parameter for the neutron decay, $A = -0.1184\,\pm\,0.0010$~\cite{tanabashi2018PDG}.
The polarized neutron beam at BL05 was used for this experiment to produce bunches by the SFC.
Although we used the polarized neutron from the SFC, there was no magnetic field to keep the polarization. Thus, the polarization direction of the neutron is unknown and 
the detection efficiency of electrons in the TPC may change due to the unexpected bias of the momentum direction of the electron.
We compared the detection efficiencies when neutrons were completely polarized along $x$-, $y$-, or $z$-axis, as well as unpolarized using the simulation.
The maximum deviation was $+0.13$\% when neutrons were polarized in the $-y$ direction, which goes to the bottom of the drift direction. The value was budgeted as an uncertainty.

It is known that $W$ value increases for low energy charged particles~\cite{blum2008particle}, although this effect was not implemented in the current simulation. 
This may be significant for protons from the neutron decay (the kinetic energy is below 1\,keV), leading to a decrease in the detection efficiency. 
The upper limit of this effect can be estimated by forcibly setting the proton kinetic energy as zero in the simulation, i.e., assuming an infinite $W$ value for the proton. 
The loss of efficiency was consistent with zero, $(0.06\,\pm\,0.35)\%$ for the neutron lifetime with an uncertainty originating from the statistical error of the simulation.

A part of the neutron decays emit not only an electron and a proton but also a $\gamma$-ray. The probability of the radiative decay is $(9.2\,\pm\,0.7)\,\times\,10^{-3} $ for $\gamma$-rays with energy of more than 0.4\,keV~\cite{bales2016precision}. 
This reaction is expected to give less effect because the TPC is insensitive to $\gamma$-rays and an electron is produced as well though its energy is reduced. 
The effect of the energy reduction was calculated using the theoretical formulation in Ref.~\cite{gaponov1996radiative}. 
The probability that the electron energy becomes less than the cut off energy (5\,keV) due to the radiative decay is expected to be 6.5$\,\times\,$10$^{-7}$, therefore we ignored this effect. 

\begin{table}[htp]
\begin{center} 
\caption{Efficiency ($\varepsilon_{\rm He}$) uncertainty budgets (Series 6)}
\begin{tabular}{ c c c} \hline
Cut name &Efficiency (\%) & Uncertainty (\%) \\\hline
$E_{\rm max}^{\rm field}$ cut ($\xi_{\rm sep}^{\rm He}$) &$-$0.01 &$^{+0.01}_{-0.00}$\\\hline\hline
$\varepsilon_{\rm He}$ & 99.99 & $^{+0.01}_{-0.00}$\\\hline
\end{tabular}
\label{table_Eff3He_Fill66}
\end{center} 
\end{table}
\begin{table}[htp]
\begin{center} 
\caption{Efficiency ($\varepsilon_{\beta}$) uncertainty budgets (Series 6)}
\begin{tabular}{ c c c} \hline
Cut name &Efficiency (\%) & Uncertainty (\%) \\\hline
$E_{\rm max}^{\rm field}$ cut ($\xi_{\rm sep}^{\beta}$) & $-$1.3 & $^{+0.5}_{-0.7}$\\
Low energy cut at $E_{\rm thresh}^{\rm anode}$ & $-$0.3 & $^{+0.1}_{-0.2}$ \\
Tritium decay rejection & $-$0.6 &  0.06 \\
Track geometry ($y$-direction) & $-$1.3 &   0.2 \\
Track geometry ($X_E$) & $-$3.2 & 0.03 \\\hline
Neutron polarization & & 0.13  \\
$W$ value for decay proton&  & 0.35  \\\hline\hline
$\varepsilon_\beta$ & 93.9 & $^{+0.6}_{-0.8}$\\\hline
\end{tabular}
\label{table_Effb_Fill66}
\end{center} 
\end{table}

\section{Result and Discussion}\label{sec_results}

From the results and discussions in the former sections, the number of events of the {\Hecapture} ($S_{\rm He}$) and neutron decay ($S_{\beta}$), the extraction efficiencies of the {\Hecapture} reactions ($\varepsilon_{\rm He}$) and neutron decay ($\varepsilon_{\beta}$), and the number density of $^3$He in the TPC ($\rho$), were obtained with uncertainties and provided in Table~\ref{table_Helium3_NoA_Fill66},~\ref{table_N3He_Fill66},~\ref{table_Nb_Fill66},~\ref{table_Eff3He_Fill66}, and \ref{table_Effb_Fill66} for a typical measurement series (Series 6), respectively. The neutron lifetime derived by Eq.~(\ref{eq_tau_0}) is listed in Table~\ref{table_All_Fill66} with all values and uncertainties. 

Note that some uncertainties are not independent; $S_{\beta}$ and $S_{\rm He}$, $\varepsilon_{\beta}$ and $\varepsilon_{\rm He}$ have negative correlations through $E_{\rm max}^{\rm field}$ cut, which bring underestimation of the uncertainty of $\tau_{\rm n}$. However, the effects were negligible (<0.1\%) because the uncertainties of $\xi_{\rm sep}^{\rm He}$ and $\varepsilon_{\rm He}$ were small enough. 
The uncertainties of $\xi_{\rm pileup}^{\beta}$ and $\xi_{\rm pileup}^{\rm He}$, which describe how the pileup events were identified, also have a negative correlation, while this effect is also less than 0.1\% and negligible.
There are more parameters which have correlations; $\xi_{\rm scat}^{\beta}$ was determined by $\xi_{\rm scat}^{\rm He}$, and the uncertainty of the energy cuts, $E_{\rm thres}^{\rm field}$ and $E_{\rm thres}^{\rm anode}$, were both determined by the discrepancy between the measurement and the simulation of the cosmic-ray. 
A part of these uncertainties cancel each other out in the estimation of the neutron lifetime. We adopted the quadratic sum of them in Table~\ref{table_All_Fill66} for a conservative and simple estimation.

\begin{table}[htp]
\begin{center} 
\caption{Values and Uncertainty budgets (Series 6)}
\begin{tabular}{ c c c c} \hline
Term & Value & Unit & Relative uncertainty(\%) \\\hline
$S_{\rm He}$ & (3.581 $\pm \,0.006\,_{\rm stat}\, ^{+0.004}_{-0.002}\,_{\rm sys })\times 10^{5}$ & events & 0.18$\,_{\rm stat}\,^{+0.11}_{-0.06}\,_{\rm sys}$\\
$S_{\beta}$ & (1.441 $\pm \, 0.039\,_{\rm stat}\, ^{+0.011}_{-0.018}\, _{\rm sys })\times 10^{4} $ & events & 2.7$\,_{\rm stat}\,^{+0.8}_{-1.3}\,_{\rm sys}\,$\\
$\varepsilon_{\rm He}$ & 99.99 $^{+0.01}_{-0.00}\,_{\rm sys}$ & $\%$ & $^{+0.01}_{-0.00}\,_{\rm sys}$\\
$\varepsilon_\beta$ & 93.9 $^{+0.6}_{-0.8}\,_{\rm sys}$ & $\%$ & $^{+0.7}_{-0.9}\,_{\rm sys}$\\
$\rho$ & 2287 $\pm 10\,_{\rm sys}$ & $10^{16}$ atoms$/{\rm m}^{3}$ & 0.4$\,_{\rm sys}$\\
$\sigma_0$ & $5333 \pm 7\,_{\rm sys}$  & $10^{28}$ ${\rm m}^{2}$ &$0.13\,_{\rm sys}$ \\
$v_0$ & 2200 & m/s & exact \\\hline\hline
$\tau_{\rm n}$ & 869 $\pm$ 24$\,_{\rm stat}$  $\, ^{+13}_{-11}\,_{\rm sys}$ & s & 2.6$\,_{\rm stat}$ $\,^{+1.5}_{-1.1}\,_{\rm sys}$\\\hline
\end{tabular}
\label{table_All_Fill66}
\end{center} 
\end{table}

For each series of the measurement, a value of the neutron lifetime with uncertainties was derived in the same manner. The results are shown in Table~\ref{table_all_lifetimes}. 
We observed no systematic effects due to the $\rho$ values described in Sec.~\ref{sec_measurement} in the present sensitivity.
The average was calculated by fitting only with statistical uncertainties, where $\chi^2$/ndf = 5.8/5. 
The systematic uncertainties of all measurement series were expected to correlate with each other. Thus, we treated them as to be fully correlated as conservative estimation; the upper and lower systematic uncertainties were determined by taking averages of the data points shifted to 1\,$\sigma$.
\begin{table}[htp]
\begin{center} 
\centering
\caption{Neutron lifetimes for each measurement series and those combined}
\begin{tabular}{c|c}
Series & $\tau_{\rm n} $ (s)             \\\hline
 1 & 951 $\pm$ 27 $\,_{\rm stat}\,^{+22}_{-34}\,_{\rm sys}$ \\ 
 2 & 906 $\pm$ 20 $\,_{\rm stat}\,^{+13}_{-12}\,_{\rm sys}$ \\ 
 3 & 908 $\pm$ 49 $\,_{\rm stat}\,^{+13}_{-34}\,_{\rm sys}$ \\ 
 4 & 890 $\pm$ 24 $\,_{\rm stat}\,^{+16}_{-15}\,_{\rm sys}$ \\ 
 5 & 882 $\pm$ 25 $\,_{\rm stat}\,^{+12}_{-19}\,_{\rm sys}$ \\ 
 6 & 869 $\pm$ 23 $\,_{\rm stat}\,^{+13}_{-11}\,_{\rm sys}$ \\\hline
Combined	&	898 	$\pm	10 \,_{\rm stat}\,^{	+15	}_{	-18 	}\,_{\rm sys}\,$	

\end{tabular}
\label{table_all_lifetimes}
\end{center} 
\end{table}
By combining all measurement series, we obtained a neutron lifetime of
\begin{equation}
\tau_{\mathrm{n}} = 898\,\pm10\,_{\rm stat}\,^{+15}_{-18}\,_{\rm sys}~\rm{s}.
\end{equation}

\begin{figure}[ht]
	\begin{center}
	\includegraphics[width=0.7\columnwidth]{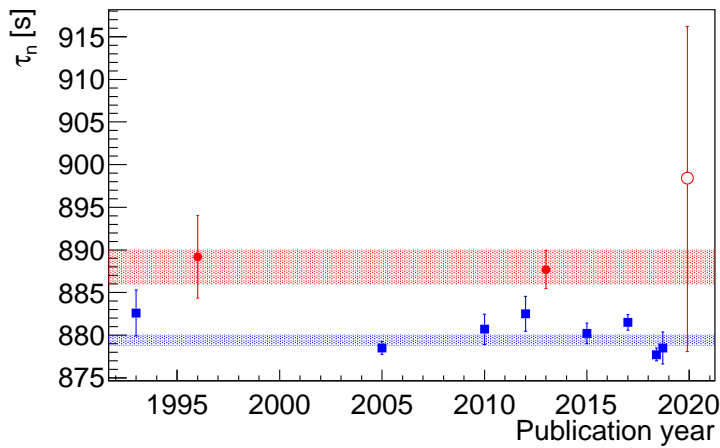}
	\end{center}
	\caption{Data of neutron lifetime obtained with the bottle method (blue square)~\cite{Mampe1993an, pichlmaier2010neutron, steyerl2012quasielastic, Arzumanov2015tea, serebrov2017new, pattie2018measurement, ezhov2018measurement} and the beam method (red circle)~\cite{Byrne1996zz,yue2013improved}. The blue and red bands show the global average and error of each method.
    The result of this work is shown by the open circle.
	}
	\label{fig_lifetimesl}
\end{figure}

By simply summing the statistic and systematic uncertainties in quadratic, it gave us 
$\tau_{\mathrm{n}} = 898\,^{+18}_{-20}\,\rm{s}$, which is shown in Fig.~\ref{fig_lifetimesl} to compare with previously published results obtained with the bottle method~\cite{Mampe1993an, pichlmaier2010neutron, steyerl2012quasielastic, Arzumanov2015tea, serebrov2017new, pattie2018measurement, ezhov2018measurement} and the beam method~\cite{Byrne1996zz,yue2013improved}. 
Both of the previously published results are within the uncertainty of our measurement. Because the uncertainty of this work is still larger than the difference between the two methods, further improvements are required to resolve the neutron lifetime puzzle by our experiment.

Improvements in the experimental accuracy are in progress for the statistics and the major systematic uncertainties mentioned below. The beam transport with larger acceptance, which is expected to increase the neutron intensity 8-fold, will be installed to improve the present statistical error of 10\,s. 
The systematic uncertainty in this work was dominated by $\xi_{\rm n \gamma}$, which corresponds to +2/-14\,s in $\tau_{\rm n}$.
Reduction or identification of the unknown background simultaneously occurred with the neutron decay, discussed in Sec.~\ref{sec:ProcD}, reduce the systematic uncertainty. Higher statistics by the new beam transport will help the identification because the present statistical error in the peripheral region was dominated by the environmental backgrounds and the increase of the beam intensity improves the signal to noise ratio. 
Additional measurements with lower pressure gas would enable us to reduce $\xi_{\rm n \gamma}$; for instance, 50 kPa operation, which is reasonably performed, makes it half of the present value.

The pileup correction $\xi_{\rm pileup}^{\beta}$ on $S_{\beta}$ shown in Table~\ref{table_Nb_Fill66} is another dominant systematic uncertainty, which corresponds to +11/-4\,s. 
This is mostly due to the cosmic-rays coming after the triggers. It can be rejected by implementing a software veto using the signals of the cosmic-ray veto counter recorded in the TDC, and it will reduce the uncertainty to +4/-0.5\,s.
The uncertainty of $\varepsilon_{\beta}$ was mainly caused by the $E_{\rm max}^{\rm field}$ cut as shown in Table~\ref{table_Effb_Fill66}, which corresponds to +4/-6\,s. 
The separation of the neutron decay and {\Hecapture} is currently performed by only one parameter ($E_{\rm max}^{\rm field}$) but it is possible to reduce the uncertainty by using another parameter of the particle trajectories, e.g. the full-waveform integration. A cut with the two parameters is expected to reduce the uncertainty to 1\,s. 

The uncertainty of $\rho$ corresponds to 4\,s in this work as shown in Table~\ref{table_Helium3_NoA_Fill66}. The main source of the uncertainty in $\rho_{\rm ad}$ was caused by the volume-ratio measurement of the vacuum chamber because the ratio was so large that it was measured in 3 steps. Employing a pressure gauge with a larger dynamic range will suppress the uncertainty of $\rho_{\rm ad}$ to 1/3 ($\sim$1\,s). The uncertainty of $\rho_{\rm G1}$ was limited by the accuracy of the mass spectroscopy. Measurements with nitrogen gas proposed in Ref.~\cite{kitahara2019_14N} can reduce it to 1/10 of the current value ($\sim$0.4\,s) in principle. With those improvements, the uncertainty of $\rho$ is expected to be reduced to $\sim$1\,s.

\section{Summary}\label{sec_summary}
The neutron lifetime puzzle, the discrepancy of 8.5\,s (4.0\,$\sigma$) between the experimental data obtained with the bottle and beam methods, is still unsolved. We have launched a new experiment using the pulsed cold neutron beam at J-PARC. In this experiment, the neutron beam was formed into bunches of $40\,$cm using a spin-flip chopper, and was injected to the TPC of 1-m length. The TPC simultaneously counted the events of the neutron $\beta$-decay and the (n,p) reaction on $^3$He whose number density was accurately controlled. The neutron lifetime was derived from the ratio of those counting rates. This experiment is classified as the beam method but dominated by the different systematic uncertainties from the previous experiments. As the first result of this experiment, we obtained the neutron lifetime of $898\,\pm\,10\,_{\rm stat}\,^{+15}_{-18}\,_{\rm sys}\,$s.
The present value of the neutron lifetime does not contradict with the other recent results within the range of its uncertainty. Further improvements in the statistical and systematic uncertainties are underway. 

\section*{Acknowledgments} 
We wish to thank Prof. S. Ikeda for his efforts to encourage the starting up of the experiment. We are grateful to Drs. M. Tanaka, Y. Igarashi, and S. Sato for helpful discussions of the DAQ circuits.
This research was supported by JSPS KAKENHI Grant Number 19GS0210, 23244047, 24654058, 26247035, 16H02194, and 19H00690. 
The neutron experiment at the Materials and Life Science Experimental Facility of the J-PARC 
was performed under user programs (Proposal No. 2012A0075, 2012B0219, 2014A0244, 2014B0271, and 2015A0316) and
S-type project of KEK (Proposal No. 2014S03 and 2019S03).

\bibliographystyle{ptephy}
\bibliography{lifetime}
%
\end{document}